\documentclass[longauth]{aa} 

\usepackage{textcomp}
\usepackage{txfonts}
\usepackage{graphicx}
\usepackage{lipsum}
\usepackage{longtable}
\usepackage{threeparttablex}
\usepackage{booktabs}
\usepackage{adjustbox}
\usepackage{rotating}
\usepackage{multirow}
\usepackage{tablefootnote}
\usepackage[colorlinks=true,citecolor=blue]{hyperref}

\newcommand{\tess}{\emph{TESS}}
\newcommand{\gaia}{\emph{Gaia}}
\newcommand{\kepler}{\emph{Kepler}}

\newcommand{\sname}{TOI-942}
\newcommand{\planetb}{TOI-942\,b}
\newcommand{\planetc}{TOI-942\,c}
\newcommand{\logrhk}{$\rm log\,R^{\prime}_\mathrm{HK}$}
\newcommand{\kms}{\,km\,s$^{-1}$} 
\newcommand{\ms}{\,m\,s$^{-1}$} 

\newcommand{\mstar}{M$_{\star}$}
\newcommand{\rstar}{R$_{\star}$}
\newcommand{\lstar}{L$_{\star}$}

\newcommand{\msun}{$M_{\odot}$}
\newcommand{\rsun}{$R_{\odot}$}
\newcommand{\lsun}{$L_{\odot}$}

\newcommand{\vsini}{$v$\,sin\,$i_\star$}   
\newcommand{\sini}{sin\,$i_\star$}

\newcommand{\teff}{$T_{\rm eff}$}

\newcommand{\mearth}{$M_{\oplus}$}
\newcommand{\rearth}{$R_{\oplus}$}
\newcommand{\prot}{$P_{\rm rot}$}

\begin{document}

   \title{The GAPS Programme at TNG XXVIII \thanks{
   Based on observations made with the Italian {\it Telescopio Nazionale Galileo} (TNG) operated by the {\it Fundaci\'on Galileo Galilei} (FGG) of the {\it Istituto Nazionale di Astrofisica} (INAF) at the {\it  Observatorio del Roque de los Muchachos} (La Palma, Canary Islands, Spain).}}

   \subtitle{A pair of hot-Neptunes orbiting the young star TOI-942}

   \author{I. Carleo\inst{1,2},
S. Desidera\inst{2},
D. Nardiello\inst{3,2},
L. Malavolta\inst{4},
A. F. Lanza\inst{5},
J. Livingston\inst{6},
D. Locci\inst{7},
F. Marzari\inst{4},
S. Messina\inst{5},
D. Turrini\inst{8},
M. Baratella\inst{4},
F. Borsa\inst{9},
V. D’Orazi\inst{2},
V. Nascimbeni\inst{2},
M. Pinamonti\inst{10},
M. Rainer \inst{11},
E. Alei\inst{12},
A. Bignamini\inst{13},
R. Gratton\inst{2},
G. Micela\inst{7},
M. Montalto \inst{4},
A. Sozzetti \inst{10},
V. Squicciarini\inst{2,4},
L. Affer\inst{7},
S. Benatti\inst{7},
K. Biazzo\inst{5},
A. S. Bonomo\inst{10},
R. Claudi \inst{2},
R. Cosentino \inst{14},
E. Covino \inst{15},
M. Damasso \inst{10},
M. Esposito \inst{16},
A. Fiorenzano\inst{14},
G. Frustagli \inst{9,17}
P. Giacobbe \inst{10},
A. Harutyunyan \inst{14},
G. Leto \inst{5},
A. Magazz{\`u} \inst{14},
A. Maggio \inst{7},
G. Mainella\inst{14},
J. Maldonado \inst{7},
M. Mallonn\inst{18},
L. Mancini \inst{19,20,10},
E. Molinari \inst{21},
M. Molinaro\inst{13},
I. Pagano \inst{5},
M. Pedani\inst{14},
G. Piotto \inst{4},
E. Poretti \inst{14,9},
S. Redfield\inst{1}
\and
G. Scandariato \inst{5}
          }

   \institute{Astronomy Department and Van Vleck Observatory, Wesleyan University, Middletown, CT 06459, USA\\ 
            \email{icarleo@wesleyan.edu}
         \and INAF -- Osservatorio Astronomico di Padova, Vicolo dell'Osservatorio 5, I-35122, Padova, Italy\\ 
         \and  Aix Marseille Univ, CNRS, CNES, LAM, Marseille, France \\ 
         \and Dipartimento di Fisica e Astronomia Galileo Galilei, Universit{\'a} di Padova, Vicolo dellOsservatorio 3, I-35122, Padova, Italy \\ 
         \and INAF - Osservatorio Astrofisico di Catania, Via S. Sofia 78, I-95123, Catania, Italy\\ 
         \and Department of Astronomy, University of Tokyo, 7-3-1 Hongo, Bunkyo-ku, Tokyo 113-0033, Japan\\ 
         \and INAF - Osservatorio Astronomico di Palermo, Piazza del Parlamento, 1, I-90134 Palermo, Italy\\ 
        \and Institute for Space Astrophysics and Planetology INAF-IAPS
Via Fosso del Cavaliere 100, 00133 Roma, Italy\\ 
        \and INAF - Osservatorio Astronomico di Brera, Via E. Bianchi 46, 23807 Merate, Italy\\ 
        \and INAF - Osservatorio Astrofisico di Torino, Via Osservatorio 20, I-10025, Pino Torinese, Italy\\ 
        \and INAF - Osservatorio Astrofisico di Arcetri, Largo Enrico Fermi 5, I-50125 Firenze, Italy\\ 
        \and ETH Z{\"u}rich, Institute for Particle Physics and Astrophysics, Wolfgang-Pauli-Str. 27, 8093 Z{\"u}rich, Switzerland\\ 
        \and INAF - Osservatorio Astronomico di Trieste, Via Tiepolo 11, 34143 Trieste, Italy\\ 
        \and Fundaci{\'o}n Galileo Galilei-INAF, 
Rambla Jos{\'e} Ana Fernandez P{\'e}rez 7, 38712 Bre{\~n}a Baja, TF, Spain \\ 
        \and INAF - Osservatorio Astronomico di Capodimonte, Salita Moiariello 16, 80131 Napoli, Italy\\ 
        \and Th\"uringer Landessternwarte Tautenburg, Sternwarte 5, D-07778 Tautenberg, Germany 
        \and Dipartimento di Fisica G. Occhialini, Università degli Studi di Milano-Bicocca, Piazza della Scienza 3, 20126 Milano, Italy 
        \and Leibniz-Institut f{\"u}r Astrophysik Potsdam (AIP), An der Sternwarte 16, D-14482 Potsdam, Germany\\ 
        \and Department of Physics, University of Rome ``Tor Vergata'', Via
della Ricerca Scientifica 1, I-00133, Rome, Italy\\ 
        \and Max Planck Institute for Astronomy, K\"{o}nigstuhl 17,
D-69117, Heidelberg, Germany\\ 
        \and INAF Osservatorio Astronomico di Cagliari \& REM, Via della Scienza, 5, I-09047 Selargius CA, Italy\\ 
 }

   \date{Received ; accepted: Nov, 25th 2020 }

 
  \abstract
   {Both young stars and multi-planet systems are primary objects that allow us to study, understand and constrain planetary formation and evolution theories.}
   {We validate the physical nature of two Neptune-type planets transiting \sname\ (TYC 5909-319-1), a previously unacknowledged young star ($50^{+30}_{-20}$ Myr) observed by the \tess\ space mission in Sector 5.}
   {Thanks to a comprehensive stellar characterization, \textit{TESS} light curve modelling and precise radial-velocity measurements, we validated the planetary nature of the TESS candidate and detect an additional transiting planet in the system on a larger orbit.}
   {From photometric and spectroscopic observations we performed an exhaustive stellar characterization and derived the main stellar parameters. \sname\ is a relatively active K2.5V star (\logrhk\ = -4.17\,$\pm$\,0.01) with rotation period P$_{\rm rot}$ = 3.39 $\pm$ 0.01 days, a projected rotation velocity \vsini=$  13.8 \pm  0.5  $ \kms and a radius of $\sim$0.9 \rsun.  We found that the inner planet, \sname\,b, has an orbital period $P_\mathrm{b}$=$4.3263 \pm 0.0011$ days, a radius  $R_\mathrm{b}$=$4.242  _{-0.313} ^{+0.376}$ \rearth\ and a mass upper limit of 16 \mearth\ at 1$\sigma$ confidence level. The outer planet, \sname\,c, has an orbital period  $P_\mathrm{c}$=$10.1605 _{-0.0053} ^{+0.0056} $ days, a radius  $R_\mathrm{c}$=$4.793 _{-0.351} ^{+0.410}$ \rearth\ and a mass upper limit of 37 \mearth\ at 1$\sigma$ confidence level. }
   {}


\titlerunning{}
\authorrunning{I. Carleo et al.}
   \maketitle
%
\section{Introduction}
\label{sec:intro}
After the \kepler\ mission \citep{Boruckietal2010} has discovered hundreds of multi-planet systems \citep{weiss2018}, enabling detailed statistical studies, now it is the turn of \tess\ (\textit{Transiting Exoplanet Survey Satellite}, \citealt{Rickeretal2014}), which is allowing to add tens of confirmed planets to the sample and $\sim$ 2,300 candidates\footnote{From \url{https://exoplanetarchive.ipac.caltech.edu/} as for October, 29th 2020}, among which several multi-planet systems \citep[e.g,][]{Huangetal2018, Quinnetal2019, Gunteretal2019, Gandolfietal2019, Crossfieldetal2019, Carleo2020b, Gilbertetal2020, Nowaketal2020}. Studying the properties of multi-planet systems, such as orbital periods, obliquities/eccentricities, planetary radii, as well as the chemistry of exoplanetary atmospheres and their hydrodynamical evolution, is essential to better constrain the planetary formation and evolution theories. 

It is now clear that many exoplanetary systems do not follow the same architecture as the Solar System. Instead, they show an extraordinary diversity, which makes it difficult to adopt a single formation scenario to all the observed systems. Multi-planet systems represent an excellent opportunity to study and compare the observable properties of the exoplanets orbiting the same star and formed under the same initial conditions. 


Planetary systems at young ages represent valuable resources to understand 
formation and migration processes, the physical evolution  of the planet themselves (e.g. gravitational contraction) and the planet evaporation under high-energy irradiation.
Up to now, TESS revealed a two-planet system HD 63433, a member of the $\sim$ 400 Myr old Ursa Major association  \citep{mann2020} and single planets around the 40-45 Myr old star DS Tuc \citep{benatti2019,newton2019}, the $\sim$20 Myr old star AU Mic \citep{plavchan2020}, and the 10-20 Myr old star 
HIP 67522 \citep{rizzuto2020}. K2 also contributed significantly to this field, with the discovery of  the youngest multi-planet system 
with transiting planets known to date \citep[the 4-planet system around the 23 Myr old star V1298 Tau, ][]{david2019} and the youngest single transiting planet
\citep[K2-33 at an age of 5-10 Myr, ][]{david2016}.
Several single and multi-planet systems were also identified in the Hyades and Praesepe open clusters  \citep[e.g., ][]{Malavoltaetal2016, rizzuto2017}.

In this paper, we report on the validation of a Neptune-sized planet and the discovery of an additional super-Neptune-type planetary companion, both transiting \sname, an active K2.5V star observed by \tess\ in Sector 5. 
With an age of $50^{+30}_{-20}$ Myr, this is the youngest multi-planet system identified by TESS so far.
The star was not previously known for being a young object, but it was selected as a promising case of a young planet-host candidate from our systematic check of stellar properties of the TESS Objects of Interest (TOI) \footnote{\url{https://tess.mit.edu/toi-releases/}}.
The presence of X-ray emission from \textit{ROSAT} and large activity from RAVE \citep{zerjal2017} alerted us on the possible
youth, which was confirmed by the detailed analysis of the \tess\ light curve and the first spectrum acquired
with HARPS-N at TNG. We then started the radial velocity (RV) follow-up in order to confirm the planet candidate, as part of the Global Architecture of Planetary Systems (GAPS) Young Objects Project \citep{Carleo2020a}. 

The paper is organised as follows. We first describe the observations of \sname, including \tess\ photometry, ground-based photometry and spectroscopy in Sections \ref{sec:tessphot}, \ref{sec:gbphot} and \ref{sec:gbspec}, respectively. We performed a comprehensive stellar characterization in Section \ref{sec:stellar}. We then presented our analysis on \tess\ photometry together with the transit fit and RV modeling in Sections \ref{sec:tess_analysis} and \ref{sec:rv}. Finally, we discuss our results
in Section \ref{sec:disc}, and draw our conclusions in Section \ref{sec:concl}.

\section{Observations and Data Reduction} \label{sec:obs}
\subsection{\tess\ photometry}
\label{sec:tessphot}
\sname\ (TYC 5909-319-1) was observed in Sector 5 of the
\tess\ mission from  Nov 15 to Dec 11, 2018 ($\sim 26.3$\,days). The
star was targeted in CCD 2 of the CAMERA 2. \sname\ was observed only
in long cadence mode (30 minutes). Identifiers, coordinates, proper
motion, magnitudes, and other fundamental parameters of \sname\ are
listed in Table \ref{tab:stellar}.

The detection of a 4-day transit signal was issued by the \tess\ Science Office QLP pipeline in Sector 5.  
The detection was then released as a planetary candidate via the TOI releases portal\footnote{\url{https://tess.mit.edu/toi-releases/}.} on 2019 July 24.
We extracted the light curve of \sname\ from the 1196 publicly
available Full Frame
Images (FFIs)\footnote{\url{https://archive.stsci.edu/tess/bulk_downloads/bulk_downloads_ffi-tp-lc-dv.html}}
by using the routine \texttt{img2lc} developed for ground-based instruments by
\citet{Nardiello2015,Nardiello2016a}, used by  \citet{Libralato2016a,Libralato2016b} and \citet{Nardiello2016b} in the case of \kepler/\textit{K2} data, and adapted to
\tess\ FFIs by \citet{Nardiello2019} for the PATHOS
project\footnote{\url{https://archive.stsci.edu/hlsp/pathos},DOI:
  10.17909/t9-es7m-vw14}. Briefly, for a target star, the routine
subtracts, from each FFI, all its neighbor sources by using empirical
Point Spread Functions (PSFs) and positions and luminosities from Gaia
DR2 catalog \citep{GaiaDR2}. After the subtraction, the routine
performs PSF-fitting and aperture photometry of the target
star. Aperture photometry is obtained with 4 different aperture radii (1-, 2-, 3-, 4-pixel).
We corrected the light curve of \sname\ by fitting it with the
Cotrending Basis Vectors extracted by \citet{Nardiello2020}. We refer
the reader to \citet{Nardiello2019,Nardiello2020} for a detailed description of the
PATHOS pipeline. In this work, we adopted the light curve obtained with the 2-pixel aperture photometry, selected on the basis of its photometric precision (r.m.s. $\sim$ 500 ppm).

\subsection{Ground-based Photometry} \label{sec:gbphot}
\subsubsection{SuperWASP}\label{sec:SW}

SuperWASP observations \citep{Butters10} of \sname\ were carried out for two consecutive seasons from September 2006 until February 2008.
From the public archive, we retrieved a total of 8307 magnitude measurements, after cleaning from outliers and removing low-quality data. The average photometric precision is $\sigma_V$ = 0.018\,mag.

\subsubsection{REM}
We observed \sname\ with the REM (Rapid Eye Mount; \citealt{Chincarini03}) 0.6\,m robotic telescope (ESO, La Silla, Chile) from December 13, 2019 to February 10, 2020 for a total of 38 nights, in the framework of the GAPS project. Observations were gathered with the ROS2 camera in the Sloan \it g$^{\prime}$r$^{\prime}$i$^{\prime}$z$^{\prime}$ \rm filters. 

We used IRAF\footnote{IRAF (Image Reduction and Analysis Facility) is distributed by the National Optical Astronomy Observatory, which is operated by the Association of Universities for Research in Astronomy (AURA) under a cooperative agreement with the National Science Foundation} and IDL\footnote{IDL (Interactive Data Language) is a registered trademark of Exelis
Visual Information Solutions.} to perform bias correction and flat-fielding of all frames, and to perform aperture photometry in order to extract magnitudes of \sname\ and of two nearby stars in the same FoV, 2MASS J05063072-2013462 and 2MASS J05062241-2012430; being not variable during our observation campaign, these were used as a comparison and check stars, respectively, to perform differential photometry of \sname.
The average photometric precision turned out to be $\sigma_g$ = 0.009\,mag and $\sigma_r$ = 0.006\,mag. However, data in the i$^{\prime}$ and z$^{\prime}$  filters turned out to be of low S/N ratio and were not suitable for the subsequent analysis.

\subsection{HARPS-N} \label{sec:gbspec}
We carried out spectroscopic follow-up observations of \sname, in the framework of the GAPS project, using HARPS-N spectrograph \citep{harpsn} mounted at Telescopio Nazionale Galileo (TNG). We acquired 33 high-resolution ($R$\,=\,115\,000) spectra  of \sname\ between September 19, 2019 and March 14, 2020, with a typical signal to noise ratio (SNR) of 30 and exposure time of 1800 seconds. The RV measurements were obtained through the offline version of HARPS-N data reduction software (DRS) available through the Yabi web application (\citealt{yabi}) installed at IA2 Data Center\footnote{\url{https://www.ia2.inaf.it}}, using the K5 mask template and choosing a width of the computation window of the cross-correlation function (CCF) equal to 80 \kms, in order to take into account the rotational broadening (\vsini $\sim$ 14 \kms, Sec. \ref{sec:vsini}). We also computed the RVs with the TERRA pipeline \citep{anglada2012}. The resulting RVs are listed in Table \ref{tab:YO38RVharpsn}. The RV dispersion results to be 141 \ms\ for DRS and 110 \ms\ for TERRA. This is due to the fact that the TERRA pipeline makes use of a template derived by an average target spectrum, that is compared with each acquired spectrum to find its RV shift rather than applying a fixed line mask to compute a CCF, which is fitted with a Gaussian as in the case of the DRS. This gives better RV measurements in the case of active stars as well as M-type dwarfs as also showed by \cite{Pergeretal2017}. We then decided to use the TERRA RVs for the analysis described in the next sections.

\section{Stellar parameters}
\label{sec:stellar}

\sname\ is a poorly studied object, with no dedicated works up to now in the literature.
Therefore, an in-depth evaluation of the stellar properties is warranted.
For this reason, we exploited the data described above and additional data from the literature, as detailed below.

\begin{table}
\centering
\caption{ Main identifiers, equatorial coordinates, proper motion, parallax, magnitudes, and fundamental parameters of \sname.}
\label{tab:stellar}
\begin{tabular}{lrr}
\hline
Parameter & Value & Source \\
\hline
\multicolumn{3}{l}{\it Main identifiers}  \\ 
\noalign{\smallskip}
\multicolumn{2}{l}{TIC}{146520535} & ExoFOP$^a$\\
\multicolumn{2}{l}{TYC}{5909-0319-1} & ExoFOP \\
\multicolumn{2}{l}{2MASS}{J05063588-2014441}  & ExoFOP \\
\multicolumn{2}{l}{\gaia}{2974906868489280768}  & \gaia\ DR2$^b$ \\
\hline
\multicolumn{3}{l}{\it Equatorial coordinates, parallax, and proper motion}  \\  
\noalign{\smallskip}
R.A. (J2000.0)	&    05$^\mathrm{h}$06$^\mathrm{m}$35.91$^\mathrm{s}$	& \gaia\ DR2 \\
Dec. (J2000.0)	& $-$20$\degr$14$\arcmin$44.21$\arcsec$	                & \gaia\ DR2 \\
$\pi$ (mas) 	& $6.5243\pm0.0295$                                    & \gaia\ DR2 \\
$\mu_\alpha$ (mas\,yr$^{-1}$) 	& $15.382 \pm 0.034$		& \gaia\ DR2 \\
$\mu_\delta$ (mas\,yr$^{-1}$) 	& $ -3.976 \pm 0.040$		& \gaia\ DR2 \\
\hline
\multicolumn{3}{l}{\it Optical and near-infrared photometry} \\  
\noalign{\smallskip}
$\tess$              & $11.046\pm0.007$     & TIC v8$^c$         \\
\noalign{\smallskip}
$G$				 & $11.6346\pm0.0016$	& \gaia\ DR2 \\
$G_{\rm BP}$   & $12.1468\pm0.0037$   & \gaia\ DR2 \\
$G_{\rm RP}$   & $10.9950\pm0.0032$   & \gaia\ DR2 \\
\noalign{\smallskip}
$B$              & $12.893 \pm 0.0.017 $         & APASS$^d$ \\ 
$V$              & $11.962 \pm 0.013$          & APASS \\ 
$V$              & $11.905 \pm 0.050$          & ASAS-SN \\ 
$B-V$            & $0.932 \pm 0.021$          & APASS \\ 
$g^\prime$       & $12.390 \pm 0.022$          & APASS \\
$r^\prime$       & $11.651 \pm 0.022$           & APASS \\
$i^\prime$       & $11.393 \pm 0.014$           & APASS \\
\noalign{\smallskip}
$J$ 			&  $10.231\pm0.022$      & 2MASS$^e$ \\
$H$				&  $9.747\pm0.024$      & 2MASS \\
$Ks$			&  $9.639\pm0.023$      & 2MASS \\
\noalign{\smallskip}
$W1$			&  $9.576\pm0.024$      & All{\it WISE}$^f$ \\
$W2$			&  $9.609\pm0.020$      & All{\it WISE} \\
$W3$             & $9.453\pm0.039$      & All{\it WISE} \\
$W4$             & $>8.478$             & All{\it WISE} \\
\hline
\multicolumn{3}{l}{\it Fundamental parameters}   \\
RV (\kms)      & $25.30 \pm 0.20$       & This work (HARPS-N) \\
RV (\kms)      & $23.68 \pm 1.10$       & \gaia\ DR2 \\
RV (\kms)      & $22.13 \pm 1.94$       & RAVE$^g$ \\ 
U\ (\kms)      & $-19.99\pm0.32$       & This work \\
V\ (\kms)      & $-19.04\pm0.28$       & This work \\
W\ (\kms)      & $0.28\pm0.26$         & This work \\
\teff\ (K) & $4969 \pm 100 $ & This work  \\  
\lstar\ (\lsun) & $0.438^{+0.036}_{-0.021}$  & This work \\   
\mstar\ ($\mathrm{M_\odot}$) & $0.880\pm0.040$ & This work \\ 
\rstar\ ($\mathrm{R_\odot}$) & $0.893^{+0.071}_{-0.053}$  & This work \\ 
Age (Myr) & $50^{+30}_{-20}$ & This work \\
$E(B-V)$ (mag)         & $0.003^{+0.014}_{-0.003}$ & This work \\
\vsini\ (\kms)         & $ 13.8\pm0.3   $       & This work \\
\prot\ (d)             & $3.39\pm0.01$       & This work \\
\logrhk\           & $ -4.17 \pm 0.01  $       & This work \\ 
$\log L_{X}$ (erg s$^{-1}$)           & $30.07   $       & This work (\textit{ROSAT}) \\
$\log L_{X}/L_{bol}$   & $-3.15  $       & This work \\ 
EW Li 6708\AA          & $ 281 \pm 5  $       & This work \\

\hline
\end{tabular}
\footnotesize{$^a$\url{https://exofop.ipac.caltech.edu/}, $^b$\cite{GaiaDR2}, $^c$\cite{Stassun2018}, $^d$\cite{Hendenetal2016}, $^e$\cite{cutri2003}, $^f$\cite{cutri2013}, $^g$\cite{rave5}} 
\end{table}




Broad band photometry as compiled from several all-sky catalogs is listed in Table \ref{tab:stellar}.
For the $V$ band magnitude we adopted the median value from ASAS-SN \citep{asas-sn}, from a time series of 291 epochs over 5 years. The photometric variability is then at least partially averaged out\footnote{\sname\, shows long term variations of about 0.05 mag over the time span of ASAS-SN observations}, as is also the case for the \gaia\, photometric results.
We estimate the interstellar reddening from interpolation of the 3D reddening maps of \citet{lallement2018}, following the procedure that will be described in Montalto et al. (in preparation). A reddening E(B-V)=$0.003^{+0.014}_{-0.003}$ has been obtained,
which is not unusual considering the distance ($\sim$ 150 pc) and galactic latitude ($\sim$ 31.8) of the target.
From the spectral energy distribution, there are no indications of the presence
of significant IR excess.

\subsection{Photometric \teff}

We obtained the photometric temperature using various color-\teff\ relationships by \citet{pecaut2013}  \footnote{Updated version available at \url{http://www.pas.rochester.edu/~emamajek/EEM_dwarf_UBVIJHK_colors_Teff.txt}, version 2019.3.22}. 
Averaging the results for $B-V$, $G_{\rm BP}-G_{\rm RP}$, $V-Ks$, $G-Ks$, and $J-Ks$, the \teff\ of \sname, and giving double weight to $B_{\rm P}-R_{\rm P}$, $V-Ks$, and $G-Ks$, because of the 
longer baseline and at least partial averaging of the photometric variability of the object,
yields \teff = 4969 K.
Similar results were obtained from \citet{2010casagrande} calibrations.
Considering calibration errors, the small scatter of the results between individual colors,  and the residual impact due of stellar variability, we adopt an errorbar of 100 K.
The spectral type corresponding to the photometric \teff\,  is close to  K2.5V, following the \citet{pecaut2013} scale.

\subsection{Spectroscopic analysis}

\sname\ is a young star, with an age close to the pre-main sequence cluster IC 2391 ($\sim$ 50 Myr), of spectral type close to K2.5V.
Moreover, \sname\ is a relatively fast rotator (\vsini = 13.8 \kms, see Section \ref{sec:vsini}). As a consequence, the number of isolated and clean lines significantly decreases, since most of them are blended with nearby features.  

It has been confirmed by different studies \citep{dorazi2009,schuler2010,2017aleo} that young ($<$100Myr) and cool (\teff$<5400$ K) stars display large discrepancies between ionised and neutral species of Fe, Ti and Cr reaching values up to +0.8 dex at decreasing \teff. Such differences alter the derivation of the atmospheric parameters, in particular the surface gravity, when derived by imposing the ionisation equilibrium. These effects could be explained with the presence of unresolved blends in the lines of the ionised species, that become more severe at decreasing temperatures \citep{2019tsantaki,2020takeda}.

The combination of both low temperature, high \vsini\ and young age prevents us from obtaining reasonable estimates of the atmospheric parameters and metallicity via the standard spectroscopic analysis through the equivalent width method. Therefore, we assume the stellar metallicity to be [Fe/H]=0.0$\pm$0.2 dex, as expected for young stars in the solar neighbourhood \citep{2013minchev}. We also adopted the photometric \teff\ in our further analysis.

\subsection{Rotation and activity}
\label{sec:rotation}

The rotation period of \sname\ was measured using the \tess\ light curve (see Sec. \ref{sec:tessphot}), ground-based photometric time series (Super WASP and REM), and the spectroscopic time series gathered with HARPS-N, as detailed below. We also characterized
the activity of the star.

\subsubsection{Rotation period from Super WASP photometric time series}

\begin{figure}
\includegraphics[width=0.72\linewidth,angle=90,trim = 0 0 0 80]{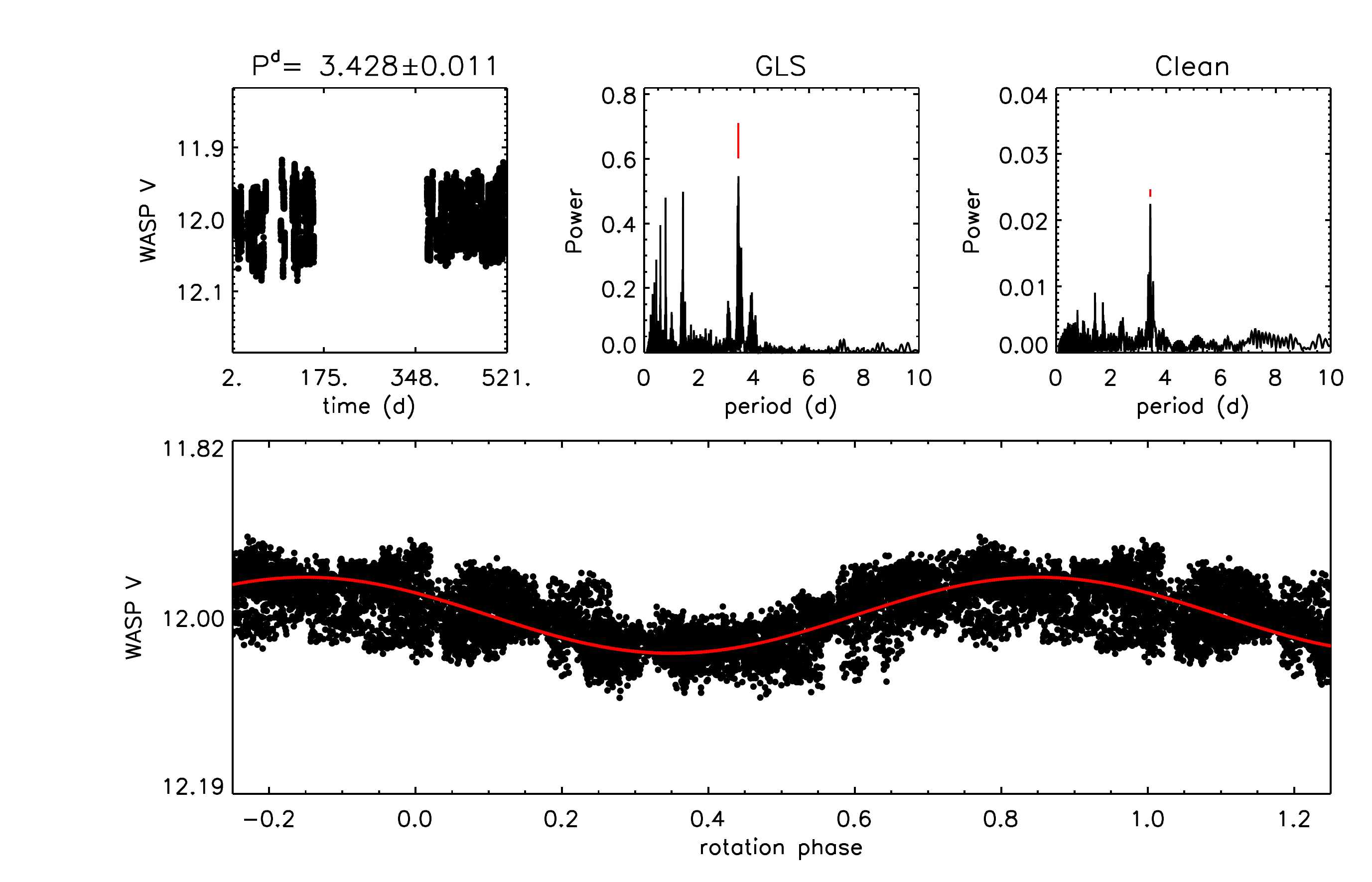}
\caption{Results of periodogram analysis of \sname. In the top-left panel, we plot the complete SuperWASP magnitudes time series vs. heliocentric Julian Day.
 In the top-middle panel, we plot the Generalized Lomb-Scargle
periodogram,
and we indicate the peak corresponding to the rotation period. In the top-right panel, we plot the CLEAN periodogram. In the bottom panel we plot
the light curve phased with the rotation period. The solid line represents the sinusoidal fit.}
\label{fig:SW}
\end{figure}

We performed a periodogram analysis of the complete data time-series and each season separately, using the Generalized Lomb-Scargle (see, e.g., \citealt{Zechmeister2009} and CLEAN \citep{Roberts87} methods. 
The Generalized Lomb-Scargle (GLS) periodogram technique makes no attempt to account
for the observational window function W($\nu$), i.e., some of the
peaks in the GLS periodogram are the result of the data sampling.
This aliasing could even account for several high peaks. The CLEAN periodogram technique tries to overcome this shortcoming by removing the effect arising from the sampling. 
We detected a rotation period $P$ = 3.428$\pm$0.011\,d with high-confidence level (False Alarm Probability FAP $<$ 0.01,  see Sect. 3.3.4) and measured  a lightcurve amplitude $\Delta$V = 0.08\,mag. Our analysis revealed a period P = 3.392$\pm$0.009\,d in the first season and P = 3.427$\pm$0.030\,d in the second season. 
FAP and uncertainty on rotation period were computed following \citet{Herbst02} and \citet{Lamm04}, respectively (see \citet{Messina10} for details).\\
In Fig.\,\ref{fig:SW}, we show a summary of our rotation period search in the case of the complete time series. 

\subsubsection{Rotation period from REM photometric time series}

\begin{figure}
   \centering
\includegraphics[width=1.0\linewidth,angle=180, ,trim = 0 0 0 30]{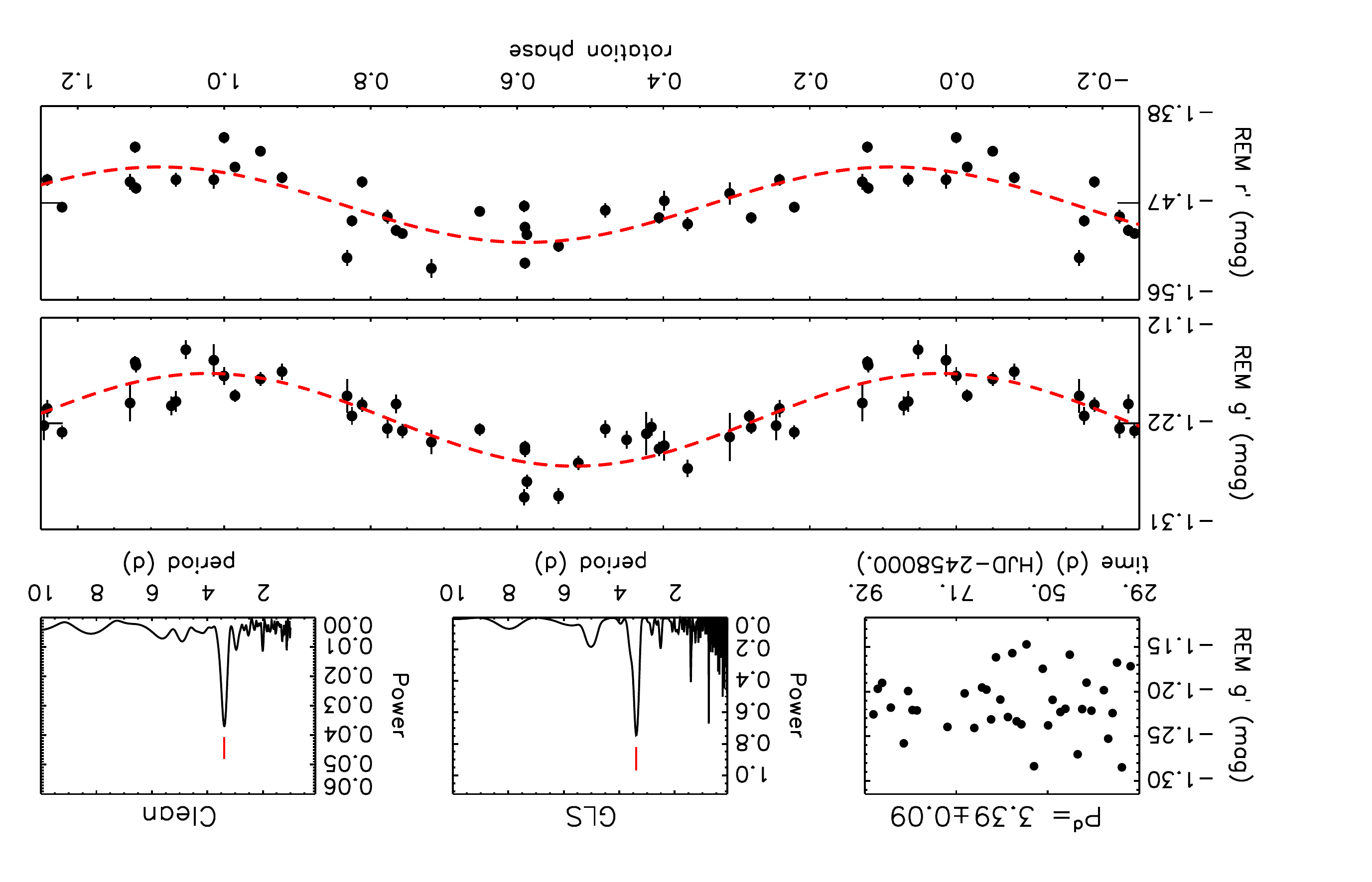}
\caption{Same as in Fig.\ref{fig:SW} but for the REM g$^{\prime}$-filter. In the bottom panel, we plot also the  r$^{\prime}$ color curve phased with the rotation period.}
\label{fig:REM}
\end{figure}

We carried out the rotation period search following the same method adopted for the SuperWASP data (see Sect.\,3.3.1) and we found a rotation period $P$ = 3.38$\pm$0.09\,d in the g$^{\prime}$-filter time series and $P$ = 3.44$\pm$0.10\,d in the r$^{\prime}$-filter time series with lightcurve amplitudes $\Delta$g$^{\prime}$ = 0.08\,mag and $\Delta$r$^{\prime}$ = 0.07\,mag (Fig. \ref{fig:REM}). Both periods are in agreement with each other within the uncertainties, and also in agreement with the period derived from SuperWASP data. The decreasing amplitude of the rotational modulation versus redder filters indicates the presence of surface temperature inhomogeneities (such as cool or hot spots) as the cause of the observed variability.

\subsubsection{Rotation period from \tess\ photometric time series}

\begin{figure}
   \centering
\includegraphics[width=1.0\linewidth,angle=-180]{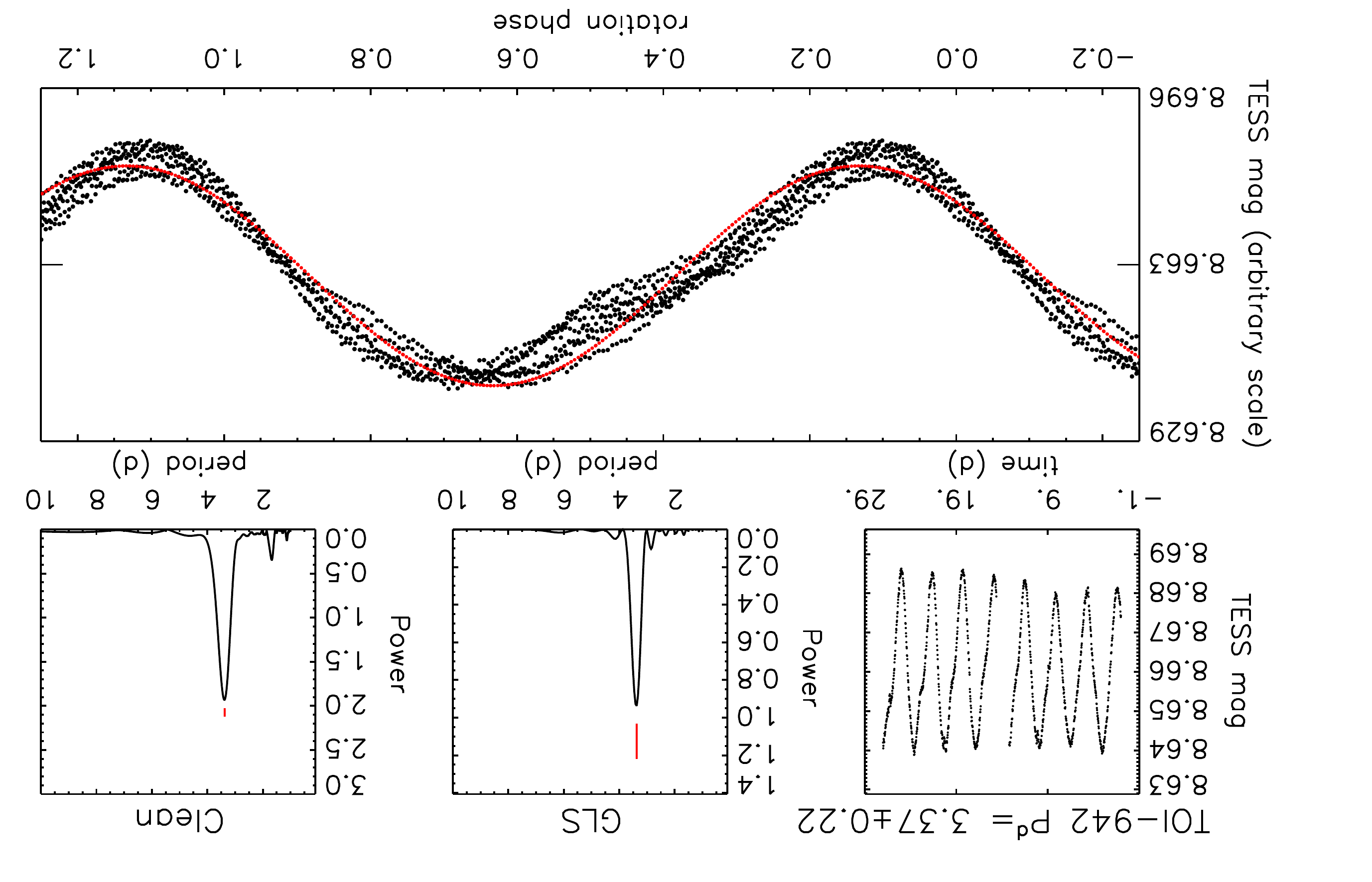}
\caption{Same as in Fig.\ref{fig:SW} but for the \tess\ timeseries.}
\label{fig:tess_lightcurve}
\end{figure}

The \tess\ photometric time series, extracted as described in Sect.\,2.1, was analysed for rotation period measurement following the same method adopted for the SuperWASP and REM data (see Sect.\,3.3.1 and Sect.\,3.3.2). The Lomb-Scargle and CLEAN analyses revealed the same rotation period  P = 3.39$\pm$0.22\,d with a very high confidence level and a $\Delta V_{\rm TESS}$ = 0.04\,mag. Despite the very high quality data, the short time base did not allow us to obtain a better uncertainty on the period measurement. In fact, the uncertainty
can be written as
\begin{equation}
    \Delta P = \frac{\delta\nu P^2}{2}
\end{equation}
where $\delta\nu$ is the finite frequency resolution of the power spectrum
and is equal to the full width at half maximum of the
main peak of the window function w($\nu$). If the time sampling
is fairly uniform, which is the case related to our observations,
then $\delta\nu$ $\simeq$ 1/T, where T is the total time span of the observations.
The results of our analysis are summarized in Fig.\,\ref{fig:tess_lightcurve}. The lightcurve shows clear evidence of the evolution of the active regions responsible for the observed rotational modulation.  The lightcurve minimum gets progressively deeper from rotation to rotation and the contribution from a secondary active region at about $\Delta\phi$ = 0.4 from the primary minimum is also evident. 

\subsubsection{Frequency analysis of the HARPS-N data and stellar activity}\label{sec:rvfreq}

\begin{figure}[t]

\resizebox{10cm}{10cm}{\includegraphics[trim={3cm 11cm 0cm 4cm},clip]{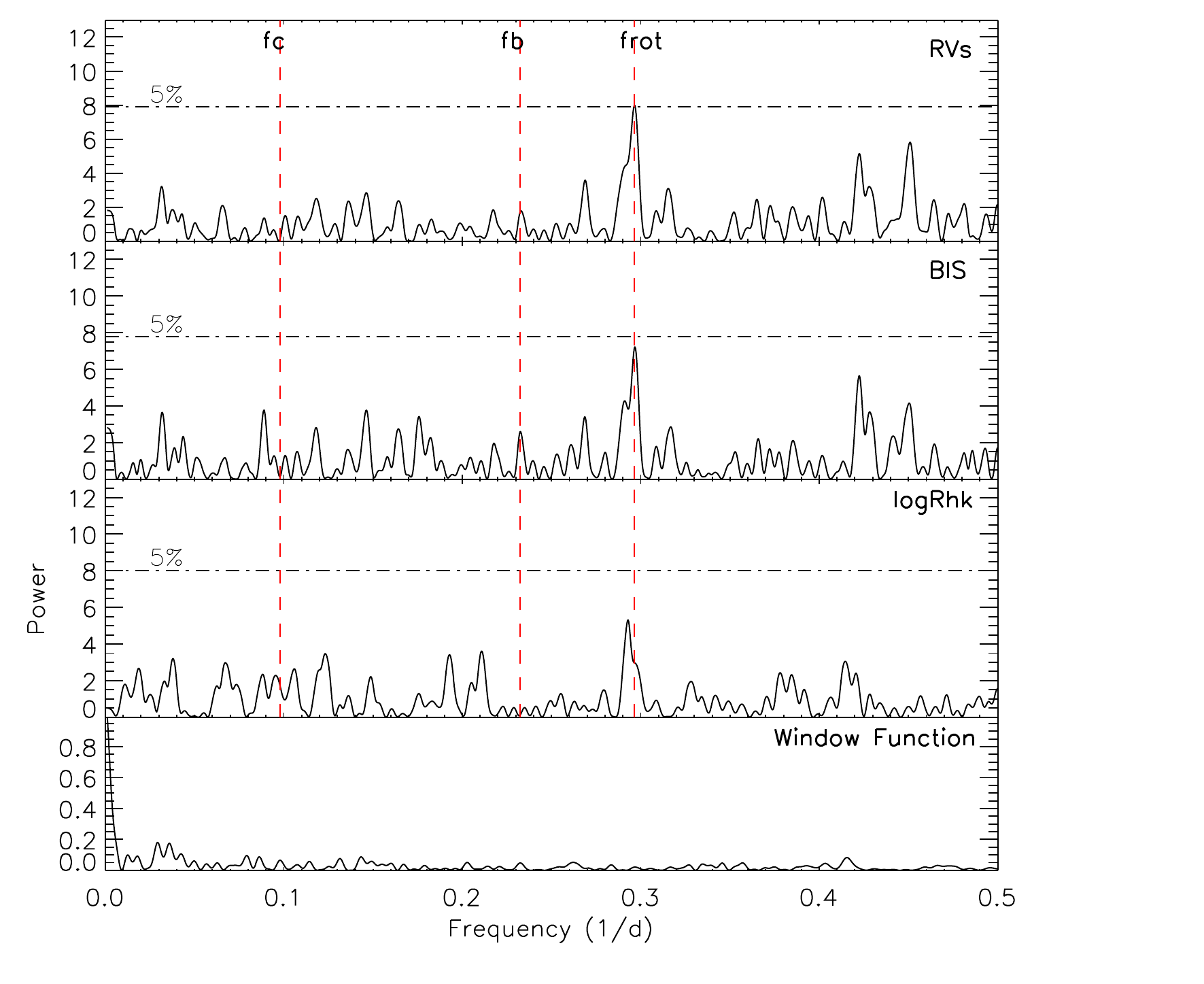}}
\caption{From top to bottom, GLS of \sname\ for HARPS-N RVs, BIS and \logrhk, and the window function. The dot-dashed horizontal lines indicate the FAP at 5\%. The vertical lines indicate the frequencies corresponding to the rotational period (f$_{rot}$), and the orbital period of the two planets (f$_{b}$ and f$_{c}$).  }
\label{fig:periodograms}
\end{figure}

\begin{figure}[t]
   \centering
\includegraphics[width=1.0\linewidth, trim = 0 13cm 0 4cm, clip]{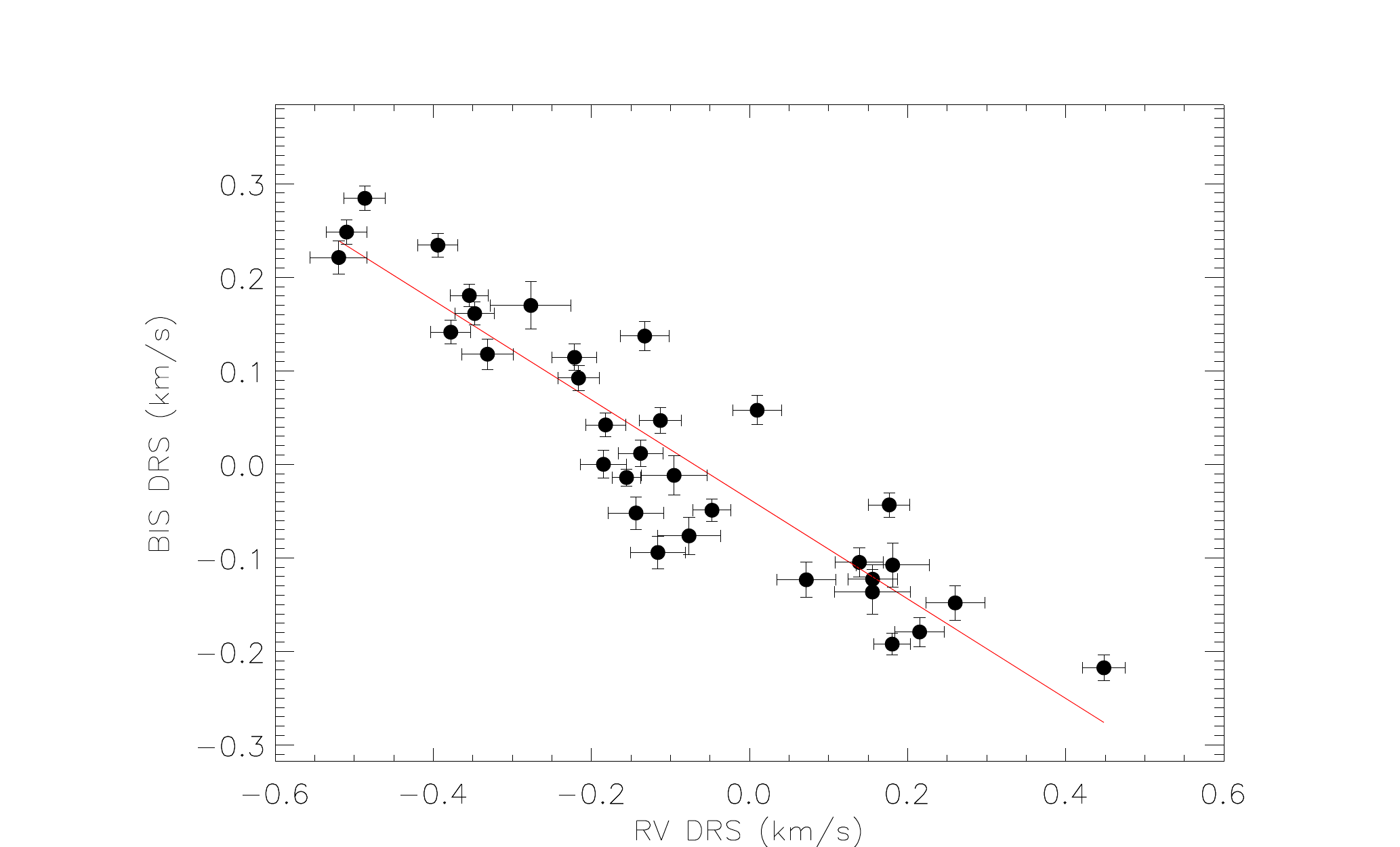}
\caption{Correlation between HARPS-N RVs and BIS of \sname.}
\label{fig:correlation_rv_bis}
\end{figure}

We performed a frequency analysis of the HARPS-N RV measurements, as well as of the Ca {\sc ii} activity index (\logrhk) and CCF asymmetry indicator (BIS). The Generalized Lomb-Scargle (GLS) periodogram \citep{Zechmeister2009} of the HARPS-N RVs shows a significant peak at 3.373 days. By performing the bootstrap method \citep{Murdoch1993,Hatzes2016}, which generates 10,000 artificial RV curves making random permutations from the real RV
values, we estimated a FAP of 4 \%; although not highly significant due to the small number of RV data points, it clearly indicates the true rotation period. Similar values of periodicity are obtained for \logrhk\ and BIS periodograms. Fig. \ref{fig:periodograms} displays the GLS for RVs, BIS and \logrhk, together with the window function. 
It is clear that the stellar activity dominates the data. This is also revealed by the strong correlation between RVs and BIS (see Fig. \ref{fig:correlation_rv_bis}), with Pearson and Spearman correlation coefficients equal to 0.92, and  a significance of 2.38$\times$10$^{-7}$, evaluated through the IDL routine {\tt R$\_$CORRELATE}.

\begin{figure}[t]
   \centering
\includegraphics[width=0.9\linewidth]{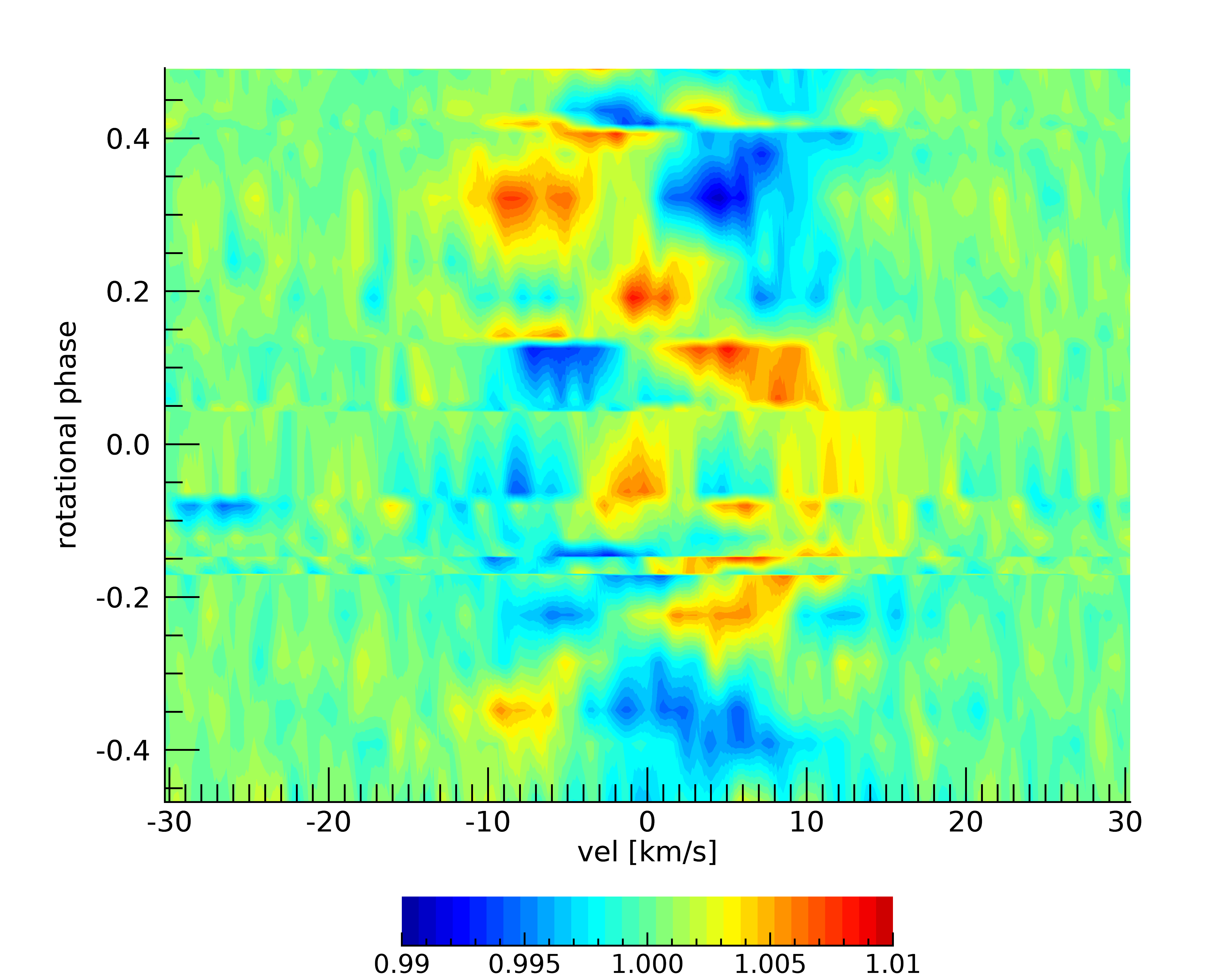}
\caption{Contour map of the CCF residuals of \sname\ versus radial velocity and rotational phase. The color bar indicates relative CCF amplitude with respect to the mean CCF.}
\label{fig:ccf_map}
\end{figure}

In order to further investigate the stellar activity, we produced the contour map of the residuals of the CCF\footnote{The CCF is provided by Yabi by comparing the spectra with a line
mask model.} versus radial velocity and rotational phase (Figure \ref{fig:ccf_map}). To obtain this map the single CCFs are subtracted from the mean CCF; positive deviations are shown in red and negative deviations are shown in  blue. The RV variation due to stellar activity can be estimated from the associated perturbation of the intensity: $\Delta RV$\,$\simeq$\,2 $\times$\,\vsini\,$\times$\,$\Delta I$\,$\times$\,$f$\,$\simeq$\,470\,$\times$\,$f$, 
where $\Delta I$\,$\sim$\,0.017 is the intensity range and $f$\,$\le$\,1 the filling factor \citep{Carleo2020a, Carleo2020b}. The contours show that the activity of the star is dominated by one main big active region, which remains quite coherent with the rotation period during the timespan of our observations.   

\subsubsection{Rotation period}

The various estimates of \prot\, derived above from photometric and spectroscopic time series agree with each other within 1$\sigma$; the small differences can be explained by 
differential rotation and evolution of active regions on the stellar surface.
We adopt a weighted mean of the various determinations, $3.39\pm0.01$ days. The comparison with other clusters and groups of known age is discussed in Sec. \ref{sec:age}.


\subsubsection{Projected rotational velocity}\label{sec:vsini}

The projected rotational velocity \vsini\ was derived in two ways. 
On one hand, we exploited the Fast Fourier Transform (FFT) method as 
in \citet{borsa2015}. This method relies on the fact that it is possible to derive the \vsini\ from the first zero positions of the Fourier transform of the line profile \citep{Dravins1990} when the rotational broadening is the dominant broadening component of the stellar line. The only prior information needed is the linear limb darkening coefficient: we adopted a value of 0.41, as found from the transit fit
in Sec. \ref{sec:tess_analysis}. We applied the FFT method on the average mean line profile and obtained \vsini\ =$13.9\pm0.3$ \kms.  

Moreover, we used a preliminary calibration of the full width half maximum (FWHM) of the CCF built from other
targets observed in the GAPS program \citep[e.g., ][]{borsa2015,bonomo2017}. We adopted the \citet{doyle2014}  relationship to take into account the contribution of the macroturbulence to the observed line width. We obtained in this way\footnote{We used, in this case, the CCF FWHM obtained with the G2 mask, as the majority
of the observed targets with similar \vsini\ are late F or G type stars.} \vsini=$13.6\pm 0.7$ \kms. 
As the two determinations agree very well, we adopt the  weighted average \vsini=$  13.8 \pm  0.5  $ \kms. 

\subsubsection{Coronal and chromospheric activity}

The mean activity level on Ca II H and K lines, as measured  with the procedure by \cite{lovis2011} adapted to the HARPS-N spectra, results in \logrhk =-4.17 $\pm$ 0.01, corresponding to 44 Myr using
\citet{mamajek2008} calibration.
The star appears also very active when using the RAVE Ca IRT index \citep{zerjal2017}, which corresponds to an age of 17 Myr with their calibration.
Finally, the star was detected in the \textit{ROSAT} all-sky survey \citep{rosatfaint}, with the X-ray source identified as 1RXS J050636.4-201439. The resulting X-ray luminosity is
large ($\log L_{X}/L_{bol}=-3.15$) and an indication of a very young star,
\citep[formally 9 Myr using ][calibration]{mamajek2008}.

\subsection{Lithium}

A very strong Lithium 6708\AA\ doublet is seen in the spectra.
We measured an equivalent width (EW) of 281 $\pm$ 5 m\AA, performing a Gaussian fit to the line profile using the IRAF task splot.
The implications in terms of stellar age are discussed in Sec. \ref{sec:age}.

\subsection{Kinematics}

The kinematics of \sname\ are fully compatible with a young star, with U, V, and W space velocities \citep[derived as in ][]{uvw} well inside the boundaries that determine the young disk population as defined by \cite{Eggen1996}. 
The star is not a member of any known moving group, as derived by the 
application of BANYAN $\Sigma$ on line tool\footnote{\url{ http://www.exoplanetes.umontreal.ca/banyan/banyansigma.php} } \citep{banyansigma}. This is not unexpected, considering the lack of members of known moving groups  in the portion of the sky where the target is located \citep[see, e.g. Fig. 5 in ][]{gagne2018}. A search for comoving objects is presented in Appendix \ref{sec:comoving}. Two objects appear to have similar kinematics and isochrone age
to \sname, and some indication of youth, and are probably comoving.

\subsection{Stellar age, radius, and mass}
Here we present the analysis aimed to obtain the stellar age, mass, radius, luminosity, as well as rotational velocity. 

\subsubsection{Stellar age}
\label{sec:age}

We compared the measurement of the age indicators for \sname\, to those of members of open clusters or groups of known age.
In this comparison, we refer \textit{i}) to the Pleiades open cluster and AB Dor moving group (MG) \citep[age 125-149 Myr]{stauffer1998,bell2015},
\textit{ii}) to the IC 2391 and IC 2602 open clusters, which have an age of 50$\pm$5 
and $46^{+6}_{-5}$ Myr, respectively, from Li depletion boundary \citep{barrado2004,dobbie2010}, 
\textit{iii}) to the Tuc-Hor, Columba, and Carina associations \citep[age 42-45 Myr, ][]{bell2015}, and the $\beta$ Pic MG  \citep[age 24-25 Myr, ][]{bell2015,messina2016}.

\begin{figure}
   \centering
\includegraphics[width=0.9\linewidth]{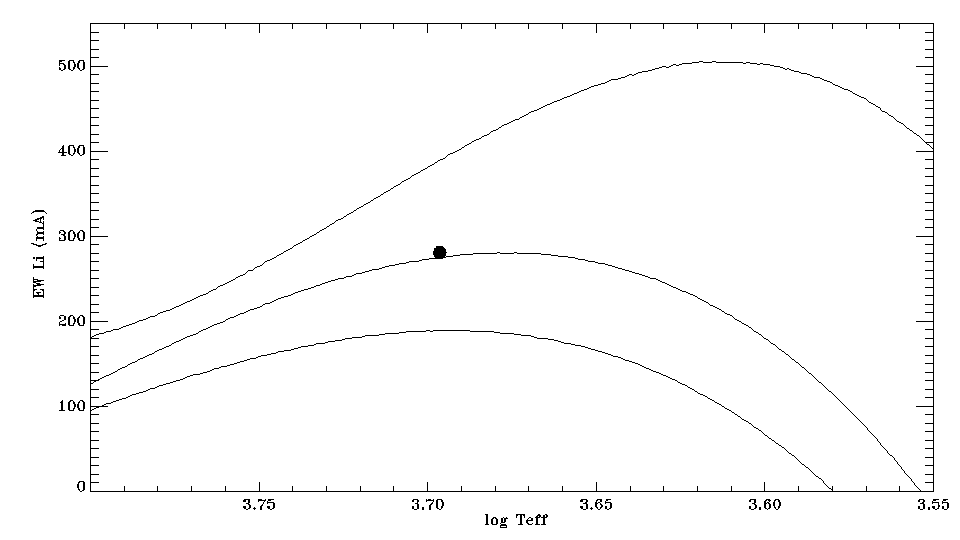} 
\caption{Lithium EW vs effective temperature for \sname\, and sequences of ScoCen, IC 2602 and Pleiades from \citet{pecaut2016}.}
\label{fig:lithium}
\end{figure}

The Li EW of \sname\ (Fig.\ref{fig:lithium}) turns out to be well above the median values of the Pleiades and AB Dor moving group (MG), although within the observed distributions \citep{desidera2015}.
The observed value is very close to the mean locus of Argus/IC 2391 \citep{desidera2011} and IC 2602 \citep{pecaut2016}
within the distribution of the members of nearby associations such
as Tuc-Hor, Columba and Carina  \citep{desidera2015},
and clearly below the locus of $\beta$ Pic MG members  \citep{messina2016}. Therefore, the age of 40-150 Myr is inferred from Lithium EW, with a most probable age close
to that of the young open clusters IC 2391 and IC2602. 

\begin{figure}
   \centering
\includegraphics[width=0.9\linewidth]{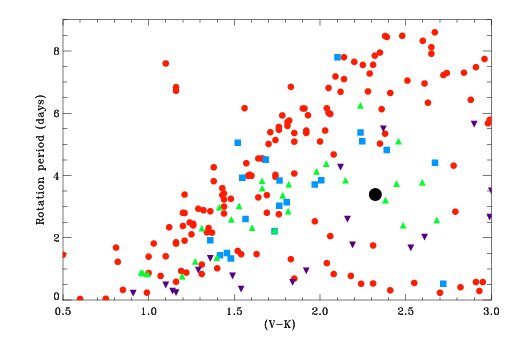}
\caption{Rotation period vs V-K (corrected for reddening for the Pleiades) for \sname\ (large black filled circle), and members of the Pleiades (red circles), IC 2391 (blue squares), Tuc-Hor, Columba and Carina association (green triangles), and $\beta$ Pic MG (purple upside-down triangles). References for rotation periods: Pleiades: \citet{rebull2016}, IC 2391: \citet{messina2011,desidera2011}, 
Tuc-Hor, Columba and Carina: Desidera et al. 2020, submitted,
\citet{Messina10,messina2011}; $\beta$ Pic MG: \citet{messina2017}.}
\label{fig:prot}
\end{figure}

Fig.\ref{fig:prot} shows the comparison of the rotation period of \sname\, with those of members of clusters and groups of known age.
The rotation period of our target is clearly faster than those of Pleiades members falling on the \it I \rm  sequence (following the \citealt{barnes2007} nomenclature), indicating a younger age,  
but slower than that of members of $\beta$ Pic MG \citep{messina2017},
confirming the older age as found for lithium. Moreover, it is  
slightly faster than the members of IC2391 open cluster and Argus association,
and more compatible with the locus of Tuc-Hor, Columba, and Carina associations (age 40-45 Myr).

The various indicators of magnetic activity are also consistent with an age of, at most, 150 Myr, although they are not able to precisely measure ages below 100 Myr. While the kinematic are fully compatible with a young age, the star is not associated with any known groups.

Finally, the position on the color-magnitude diagram (CMD) is slightly above the standard main sequence by \citet{pecaut2013} \footnote{Updated version available at \url{http://www.pas.rochester.edu/~emamajek/EEM_dwarf_UBVIJHK_colors_Teff.txt}, version 2019.3.22}, indicating a pre-main sequence status,  and is close to the sequence of the single stars bonafide members of Tuc-Hor, Columba and Carina from \citet{banyansigma} (Fig. \ref{fig:cmd}). The position on CMD of the seven possible comoving objects presented in Appendix \ref{sec:comoving} indicates a variety of ages. Two of them are close to the Tuc-Hor sequence and are promising candidates for being coeval objects truly associated
kinematically to \sname. 

\begin{figure}
   \centering
\includegraphics[width=0.9\linewidth]{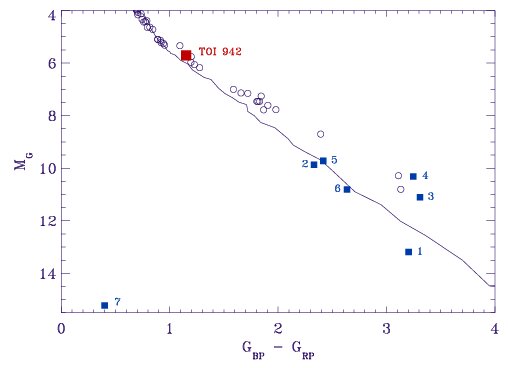}
\caption{Color-magnitude diagram of \sname\ (red large filled square) and of the possible comoving objects 
 (blue filled squares, with the number corresponding to the identification in Appendix \ref{sec:comoving}). Overplotted the main sequence locus (continuous lines) from \citet[][ updated version from the web site]{pecaut2013}, and the data of bonafide members from \citet{banyansigma} of Tuc-Hor, Columba and Carina associations \citep[age 42-45 Myr][]{bell2015}, plotted as open circles.}
\label{fig:cmd}
\end{figure}

From the findings described above, it results that the  position on the CMD and the 
results of indirect methods such as lithium and rotation nicely agree on an age close
to that of Tuc-Hor association and of IC2391 and IC2606 open clusters (45, 50, and 46 Myr, respectively, see above). Ages as young as $\beta$ Pic MG (24-25 Myr) and as old as Pleiades
and AB Dor MG (125-149 Myr) are excluded by the data. We then adopt an age of 50 Myr, with an age range of 30-80 Myr.

\subsubsection{Stellar mass, radius, and luminosity}
\label{sec:massradius}

From the adopted \teff\ and the corresponding bolometric
corrections from the \cite{pecaut2013} tables, we infer a stellar
luminosity of $0.438^{+0.036}_{-0.021}$ \lsun\
and a stellar radius of  $0.893^{+0.071}_{-0.053}$  \rsun.
The stellar mass derived through the PARAM web interface
\citep{param}\footnote{\url{http://stev.oapd.inaf.it/cgi-bin/param_1.3}}, isolating the age range allowed for the target, results
of $0.880\pm0.031$ \msun.

As a sanity check, we derived the stellar parameters with the EXOFASTv2 tool \citep{Eastmanetal2019} 
by fitting the stellar Spectral Energy Distribution (SED) and using the MIST stellar evolutionary tracks \citep{Dotter2016}. 
For the SED we considered the WISE mid-IR $W1$, $W2$ and $W3$ magnitudes \citep{cutri2013}, the 2MASS near-IR $J$, $H$ and $Ks$ magnitudes \citep{cutri2003},  
and the optical APASS Johnson $B$ and $V$ magnitudes, and Sloan $g'$, $r'$, and $i'$ \citep{Hendenetal2016}.
We imposed a Gaussian prior on the Gaia parallax and uninformative priors on all the other parameters with upper bounds of 200 Myr and 0.050 on the stellar age and 
V-band extinction $A_V$, respectively. 
We found $R_\star=0.9286 \pm 0.0087~\rm R_\odot$, $M_\star=0.912 \pm 0.032~\rm M_\odot$, 
$L_\star= 0.416 \pm 0.006~\rm L_\odot$, $\rho_\star=1.605 \pm 0.074$\,g\,cm$^{-3}$, $T_{\rm eff}=4810 \pm 23$~K, $\rm [Fe/H]=0.29^{+0.13}_{-0.16}$ dex, 
and a fairly precise age of $34 \pm 6$~Myr. The EXOFASTv2 analysis would thus indicate a possibly higher metallicity, though consistent with zero within $2\sigma$, a
slightly lower $T_{\rm eff}$ and younger age. 
Nonetheless, the stellar mass, radius, and age from EXOFASTv2 are fully consistent with the values that were independently derived above, i.e. 
$R_\star=0.893^{+0.071}_{-0.053}~\rm R_\odot$, $M_\star=0.88 \pm 0.031~\rm M_\odot$, and age of $50_{-20}^{+30}$~Myr, which we adopt as the final stellar parameters 
for the more conservative uncertainties on the stellar radius and age. 

In order to check this model-dependent result, we considered the 
dynamical masses derived for three objects of similar spectral type and
comparable age, namely the components of the system HII2147 in the Pleiades open cluster \citep{torres2020} and AB Dor A in the AB Dor moving group
\citep{azulay2017}. Both AB Dor A and HII 2147B have G band absolute magnitude
slightly fainter than \sname\ (5.75 and 5.8 vs 5.71, respectively), with 
BP-RP color slightly bluer (1.10 and 1.08 vs 1.15, respectively, with the difference likely due to the slightly younger age of \sname). 
Their dynamical masses are $0.90\pm0.08$ \msun\ for AB Dor A and $0.879\pm0.022$ \msun\
for HII 2147B. The slightly brighter primary component of the HII2147 system
(MG=5.25, BP-RP=0.97) has a dynamical mass of 0.978$\pm$0.024 \msun.
We then conclude that the mass derived from models for \sname\ is consistent with the available empirical dynamical masses of stars of comparable age. 
We then adopt the mass derived above, conservatively increasing the errorbar to 0.04 \msun\ to take the systematic uncertainties of the models into account.

\subsubsection{System inclination}

Coupling the radius and the rotation period we obtain a rotational velocity of 13.3 \kms, slightly smaller but very close to the observed \vsini. The nominal parameters yield \sini\ slightly larger than unity. From the adopted errorbars we obtain \sini\ = 1.04$_{-0.10} ^{+0.09}$. Considering only physical values we then have $i$ $\textgreater$ 70 deg. 
An alignment between the stellar equator and the orbits of the transiting planets  is then very likely.

\section{\tess\ photometric analysis and planet detection}
\label{sec:tess_analysis}
A preliminary analysis of the \tess\ data allowed us to notice an additional signal in the light curve with a period of 10 days, associated with a second transiting planet.

\begin{figure}
   \centering
\includegraphics[width=1.4\linewidth,trim = 0cm 16cm 0cm 2cm]{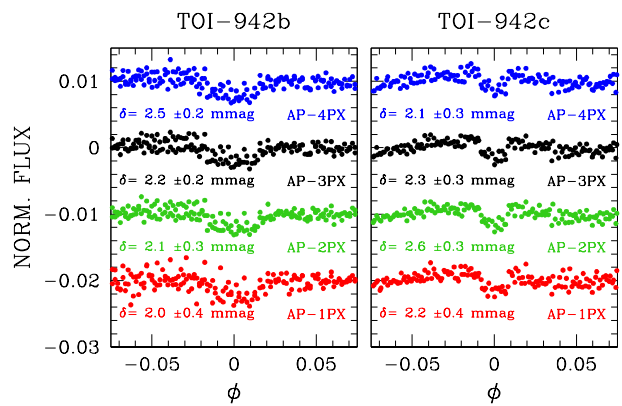}
\caption{Phased light curves for \sname\,b (left panel) and \sname\,c (right panel) obtained with different apertures marked with different colors.}
\label{fig:diff_aper}
\end{figure}
\begin{figure}
   \centering
\includegraphics[width=1.1\linewidth,trim = 0cm 8cm 0cm 2cm]{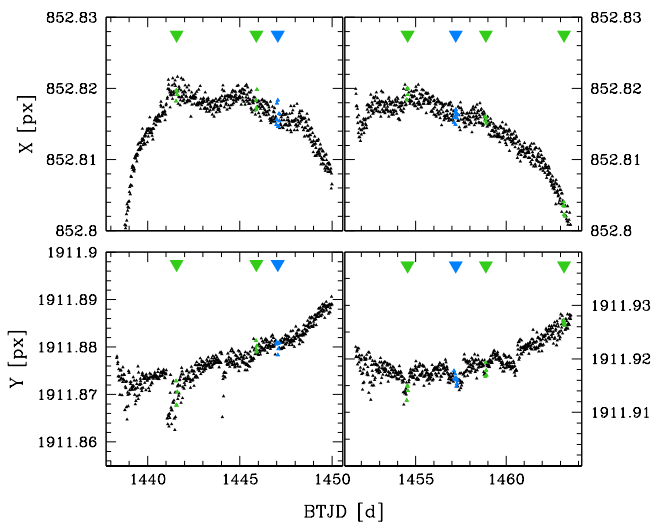}
\caption{Variation among the time of X and Y positions in pixel. Green triangles indicate the \sname\,b transits, while blue triangle the \sname\,c transits. }
\label{fig:xypos}
\end{figure}
\begin{figure*}
    \centering
    \includegraphics[bb=0 300 576 693, width=0.9\textwidth]{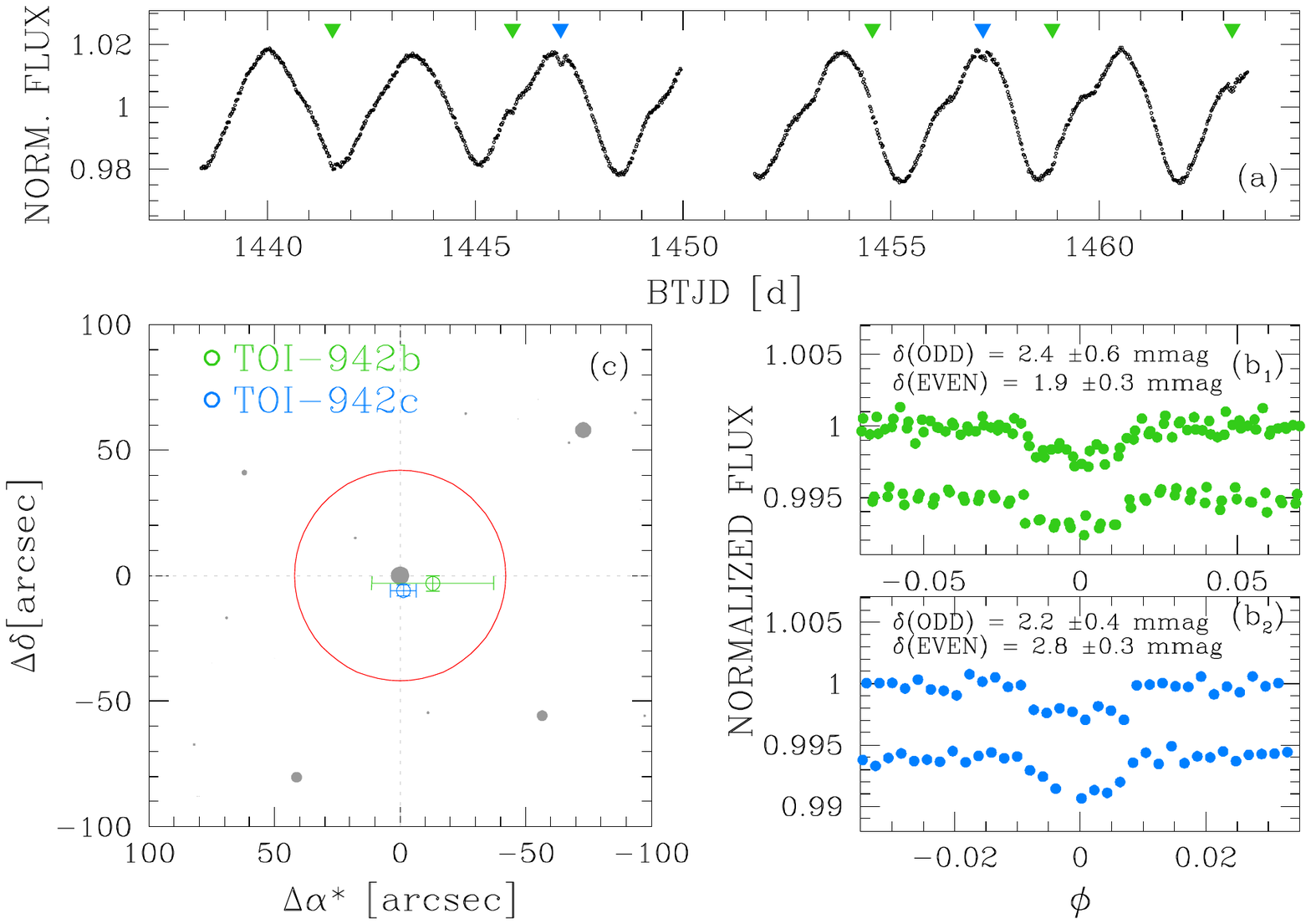}
    \caption{Vetting procedure for \sname\,b (green) and \sname\,c (blue). Panel (a) shows the normalized light curve of \sname: green and blue arrows mark the position of the single transits of \sname\,b and \sname\,c, respectively. Panels (b) show the comparison between the depths $\delta$ of the odd and even transits for the two exoplanets. Panel (c) is a finding chart, centred on \sname\ and based on the Gaia DR2 catalog: green and blue points represent the centroids computed analysing the image obtained from the difference between the out- and in-of-transit stacked images; the red circle is the photometric aperture adopted in this work (see text and \citealt{Nardiello2020} for details).  }
    \label{fig:vetting}
\end{figure*}

In order to verify that \sname\,b and \sname\,c are genuine transiting candidates, we performed three different tests on the TESS light curve (see Sec.~3.1 of \citealt{Nardiello2020} for a detailed description of the vetting tests). First, we verified that, considering light curves obtained with different photometric methods, the depth of each single transit does not change. In Fig.~\ref{fig:diff_aper} we compare the phased light curves centred for \sname\,b (left panel) and \sname\,c (right panel) obtained with different photometric apertures: for both the planets, the shape and the depth of the transits obtained with different apertures are in agreement within $1\sigma$. As a second test, we checked if flux drops due to the transits and (X,Y)-positions obtained by PSF-fitting are correlated (Fig.~\ref{fig:xypos}): there is no clear correlation between the two quantities. The third test consists in the comparison between the depths of odd and even transits, in order to exclude the possibility that the transits are due to a close eclipsing binary with different components. In panel (a) of Fig.~\ref{fig:vetting} we marked the position of the five transits of \sname\,b (green) and two of \sname\,c (blue). Panels (b1) and (b2) show the comparison between the average depths of odd and even transits for \sname\,b and \sname\,c, respectively: the odd/even transit depths are in agreement within 1$\sigma$. Finally, we computed the in/out-of-transit difference centroid for the two transit signals, in order to check if the transits are due to a contaminant. As described in \citet{Nardiello2020}, we calculated the centroid in a region of $10 \times 10$ {\it TESS} pixels ($\sim 210 \times 210$ arcsec$^2$) centred on \sname\ as follows: we selected the FFIs corresponding to the in-of-transits and out-of-transits points of the light curve and, for each transit, we calculated the stacked out-of-transit and in-of-transit image, and the difference between the two stacked images. For each transit, we calculated the photocenter on the out-/in-of transit difference stacked image and its offset relative to the Gaia~DR2 position of \sname. Finally, for each planet, we calculated the final in/out-of-transit difference centroid as the mean of the offsets associated with the single transits. Panel (c) of Fig.~\ref{fig:vetting} shows the results for the two exoplanets: in both cases, the in/out-of-transit difference centroid to the position of \sname\ is within the errors.


The transit fit was performed using the package {\tt PyORBIT}\footnote{Available at \url{https://github.com/LucaMalavolta/PyORBIT}} (\citealt{Malavoltaetal2016,Malavoltaetal2018}), a package for modelling planetary transits and radial velocities while taking into account the effects of stellar activity and astrophysical contaminants. The transit modelling relies on the popular package \texttt{batman} \citep{Kreidberg2015}. 

We modeled the \tess\ light curve with a two-planets model (ecc2p), which includes the time of first transit $T_c$, the orbital period $P$, the eccentricity $e$ and argument of periastron $\omega$ following the parametrisation from \citealt{Eastmanetal2013} ($\sqrt{e}\cos\omega$,$\sqrt{e}\sin\omega$), the limb darkening (LD) following \cite{Kipping2013}, the impact parameter $b$, the scaled planetary radius $R_{P}/R_{\star}$. For each transit, the modulation induced by stellar activity is modelled by fitting a 3rd degree polynomial (in order to take into account the variability of the light curve over a few hours) on the out-of-transit part of the light curve around each transit event. A jitter term is included 
in order to take into account possible \tess\ systematics and short-term stellar activity noise. 
We implemented a Gaussian prior on the stellar density using the stellar mass and radius provided in Section \ref{sec:stellar}. We made use of the parametrisation where the impact parameter $b$ and the stellar density $\rho_{\star}$ are free parameters (e.g. \citealt{Frustaglietal2020}).
Because no (bright) contaminants fall inside the photometric aperture adopted (red circle in panel (c) of Fig.~\ref{fig:vetting}), the dilution factor is negligible and it is not included in the fit. 

Modelling 7 transits and taking into account all the parameters described above, the number of free parameters of our model is 50. We ran the sampler for 100,000 steps, with 200 walkers, a burn-in cut of 20,000 steps, and a thinning factor of 100. In this way, we obtained 147,200 independent samples. The posteriors confidence interval was computed by taking the 34.135th percentile from the median. 

We also performed a transit fit with a circular model (circ2p). We computed the Bayesian Information Criterion (BIC) and the  Akaike  Information  Criterion  (AICc;  corrected  for  small sample sizes), which is a second-order estimator of information loss, in order to assess the quality of our fits. We obtained that the circular fit is slightly preferred to the eccentric one (see Table \ref{tab:rvmodelcomp}), but it leads to a stellar density of 0.82 $\pm$ 0.09 $\rho_{\odot}$. This value would require a full reshaping of the stellar parameters, inconsistent within the error bars. For this reason, we decided to adopt the eccentric fit to model our data, which brings to a stellar density consistent with the one obtained from spectroscopy (Sec. \ref{sec:stellar}).

\begin{table}
\centering
\caption{Comparison between transit and RV models. The model name, AICc, BIC values and number of free parameters are listed. \label{tab:rvmodelcomp}}
\begin{tabular}{lccc}
\hline
Transit Model &     BIC & AICc  & $N_\mathrm{free}$  \\
\hline
circ2p  &  -2777  &  -2915  &    46   \\
ecc2p & -2755   &  -2905   &   50    \\
\hline
RV Model &     AICc & BIC & $N_\mathrm{free}$  \\
\hline
GP only & 463 & 457 & 6 \\
circ2p+GP & 463 & 461 & 12\\
ecc2p+GP &  471 & 482    &     16  \\
ecc3p+GP &  519 & 572    &     21  \\
ecc2p+GP+trend &  502 & 512    &     17  \\
\hline
\end{tabular}
\end{table}

\begin{figure}
   \centering
\includegraphics[width=0.9\linewidth]{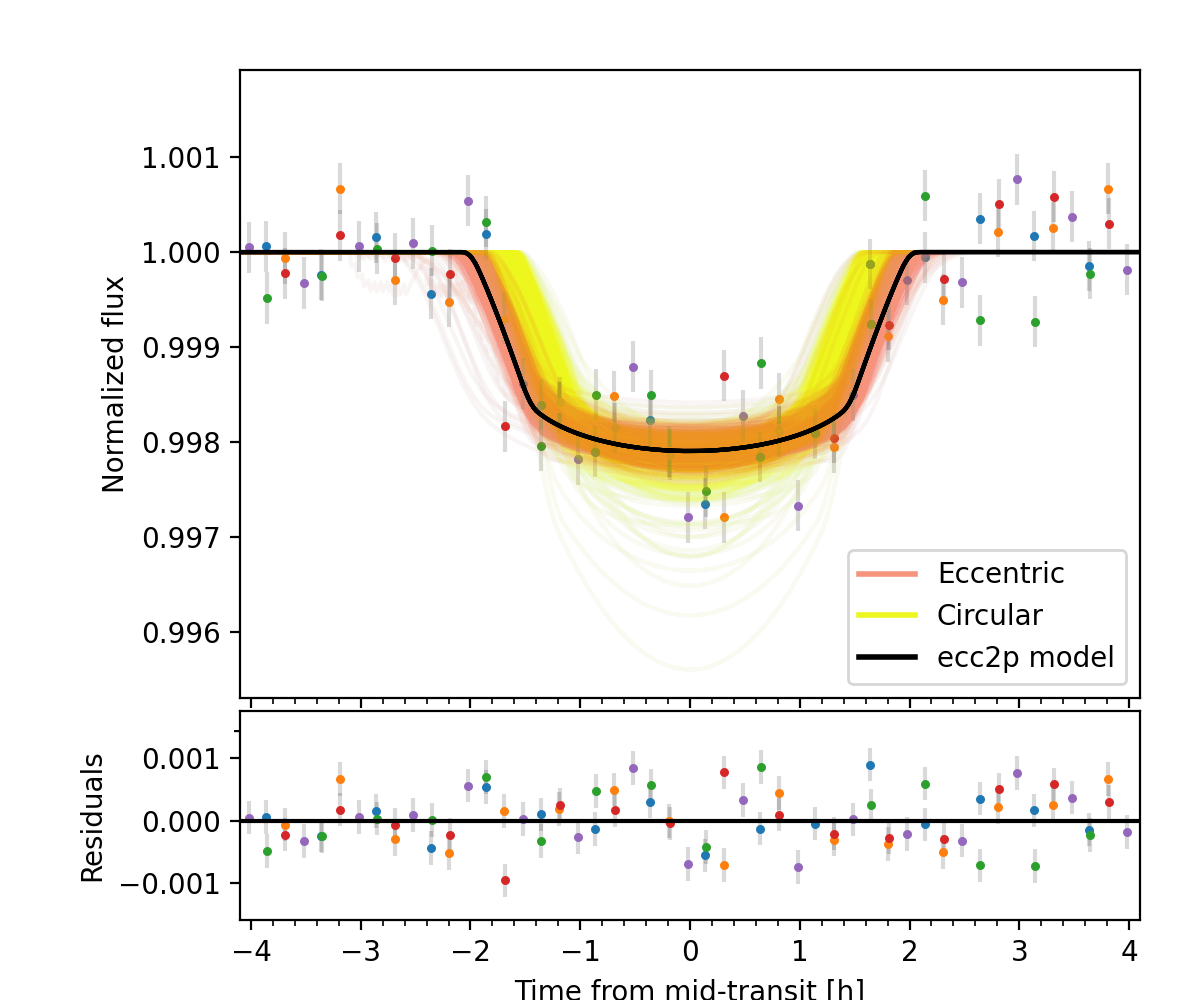}
\includegraphics[width=0.9\linewidth]{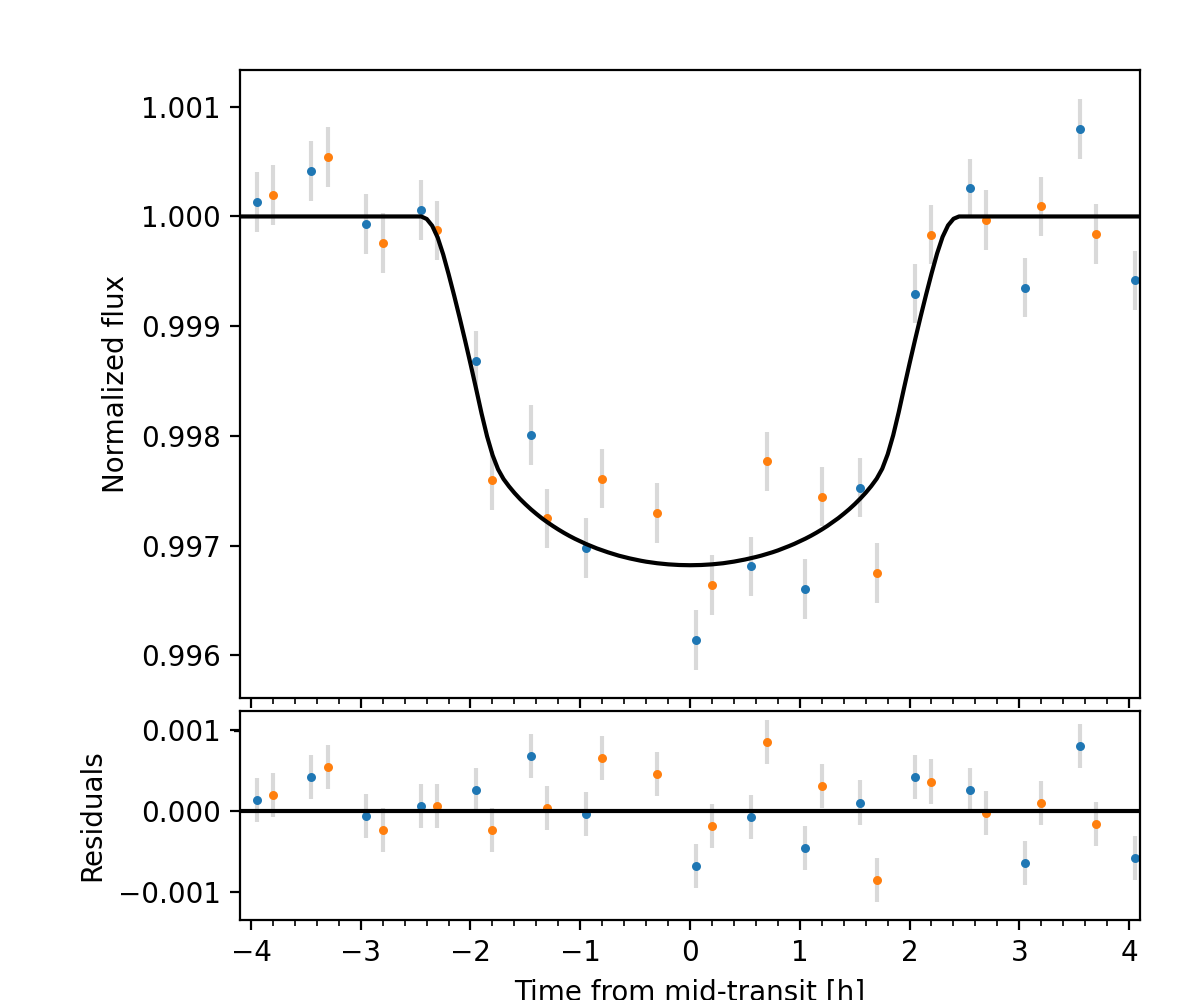}
\caption{\textit{Upper panel}: \tess\ light curve around the transit with residuals of \sname\,b. The black fit is the inferred ecc2p transit model, while orange and yellow fits represent the eccentric and circular models, respectively, obtained randomly varying all the orbital parameters. Different dot colors indicate the 5 different transits for \sname\,b.  \textit{Bottom panel}: \tess\ light curve around the 2 transits of \sname\,c, with the ecc2p model overplotted.}
\label{fig:transits}
\end{figure}

The \tess\ light curves, together with the resulting fits from ecc2p model, are shown in Figure \ref{fig:transits}. In addition, to visualize how the transit fit can change on varying the transit parameters, we simulated several models for planet b in case of circular and eccentric orbit and over-plotted them to the nominal ecc2p model fit. 

\begin{figure}
   \centering
\includegraphics[width=1.0\linewidth, trim = 0 13cm 0 4cm, clip]{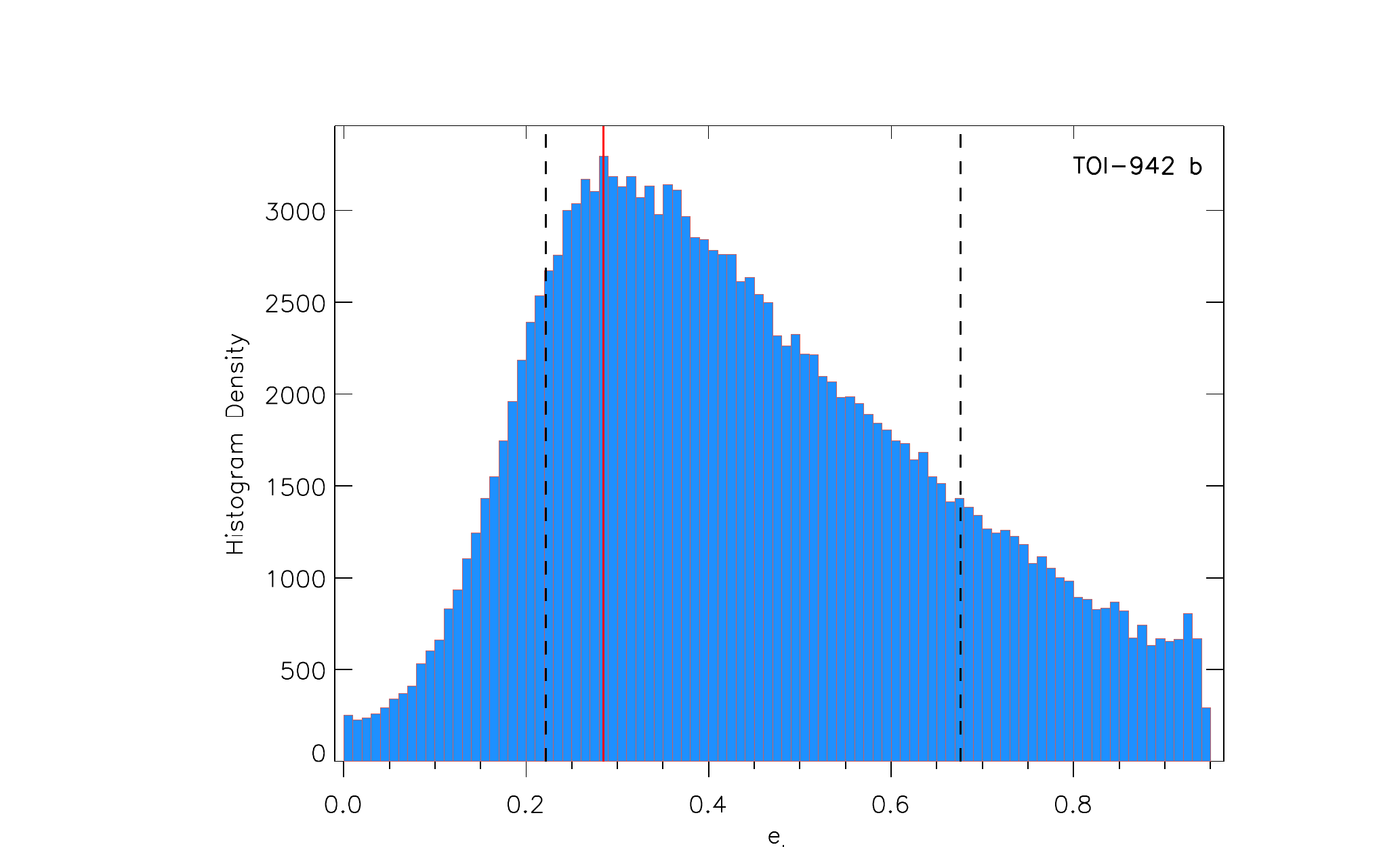}
\includegraphics[width=01.0\linewidth, trim = 0 13cm 0 4cm, clip]{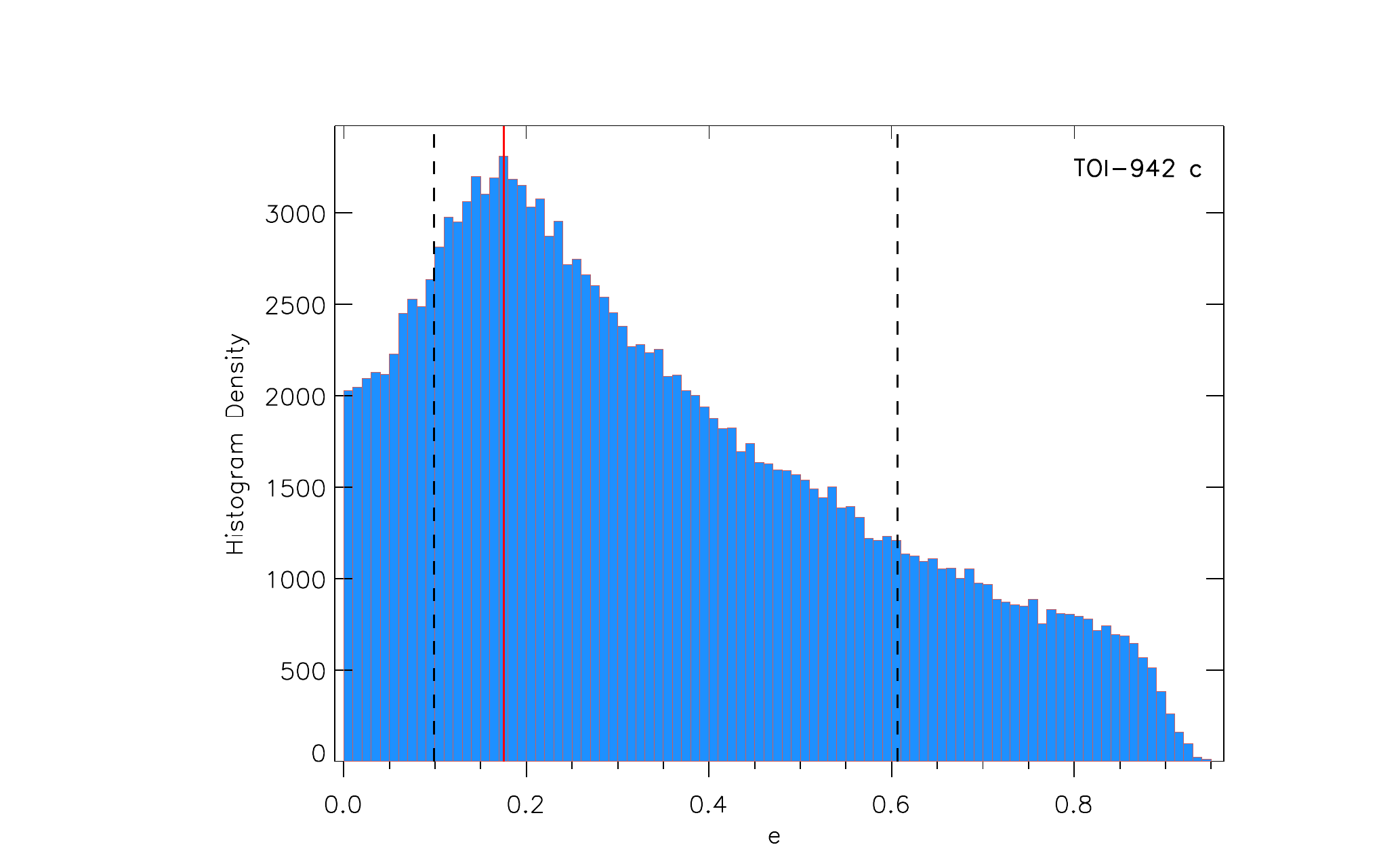}
\caption{The posterior distribution for the eccentricity of \sname\,b (upper panel) and \sname\,c (lower panel). The vertical red line indicates the maximum value of the distribution, while the dashed lines indicate the 16-th and 84-th percentiles.}
\label{fig:ecc}
\end{figure}

The fitted parameters, the adopted priors and the parameters estimates obtained from the eccentric model are listed in Table \ref{tab:transit_params}. We found that the inner planet (\sname\,b)  has an orbital period of $P_{\mathrm{b}}$ = $4.3263 \pm 0.0011$ days, and a radius  $R_{\mathrm{b}}$ = $ 4.242  _{-0.313} ^{+0.376} $ \rearth, while the outer planet (\sname\,c) has an orbital period of $P_{\mathrm{c}}$ = $10.1605 _{-0.0053} ^{+0.0056} $ days, and a radius  $R_{\mathrm{c}}$ = $ 4.793 _{-0.351} ^{+0.410}$ \rearth. We noticed that the periastron argument for both planets is $\sim$268 deg; this is mainly due to a numerical bias, which leads to this configuration while minimizing the eccentricities and maximizing the transit duration. Another interesting aspect is the slightly eccentric orbit for \sname\,b. While the transit data do not put significant constraints on the eccentricities, the transit duration, which is related to the stellar density, imposes a lower limit. Figure \ref{fig:ecc} shows the posterior distributions for the eccentricity of both planets. Being both quite broad, we decided to adopt the peak values, which correspond to 0.285$_{-0.099} ^{+0.133}$ for \sname\,b and 0.175$_{-0.103} ^{+0.139}$ for \sname\,c. We also found that planet b and c have eccentricities $\leq$ 0.05 at 0.8\% and at 8\% cumulative percentages, respectively. We further discuss the eccentricity issue in Sec. \ref{sec:disc}. 

\begin{table*}
  \footnotesize
  \caption{\sname\ parameters from the transit and RV fits. \label{tab:transit_params}}  
  \centering
  \begin{tabular}{lcc}
  \hline
  \multicolumn{3}{c}{\textbf{Transit fit}} \\
  \hline
  Parameter & Prior$^{(\mathrm{a})}$ & Value$^{(\mathrm{b})}$  \\
  \hline
  \multicolumn{3}{l}{\emph{ \bf Model Parameters for \sname b}} \\
    Orbital period $P_{\mathrm{orb}}$ (days)  & $\mathcal{U}[4.3, 4.5]$   & $ 4.3263 \pm 0.0011 $ \\
    Transit epoch $T_0$ (BJD - 2,450,000)  & $\mathcal{U}[8441.40, 8441.70]$  & $ 8441.571389 _{-0.003565 } ^ { +0.003668 }$ \\
    $\sqrt{e} \sin \omega_\star$ &  $\mathcal{U}(-1,1)$ & $ -0.501 _{-0.151} ^ {+0.131} $ \\
    $\sqrt{e} \cos \omega_\star$  &  $\mathcal{U}(-1,1)$ & $-0.013 _ {-0.475} ^ {+0.485}$ \\
    Scaled planetary radius $R_\mathrm{p}/R_{\star}$ &  $\mathcal{U}[0,0.5]$ & $0.0425 \pm 0.002$  \\
    Impact parameter, $b$ &  $\mathcal{U}[0,1]$  & $ 0.309 _{-0.216} ^ {+0.292} $ \\
    %
    \multicolumn{3}{l}{\emph{ \bf Model Parameters for  \sname c}} \\
    Orbital period $P_{\mathrm{orb}}$ (days)  &  $\mathcal{U}[10.0, 10.3]$ & $ 10.1605 _{-0.0053} ^{+0.0056} $ \\
    Transit epoch $T_0$ (BJD - 2,450,000)  & $\mathcal{U}[8446.90, 8447.20]$  & $ 8447.054230 _{-0.004119} ^{+0.003941}$  \\
    $\sqrt{e} \sin \omega_\star$ &  $\mathcal{U}(-1,1)$ &  $-0.358 _ {-0.185} ^ {+0.196}$\\
    $\sqrt{e} \cos \omega_\star$  &  $\mathcal{U}(-1,1)$ & $-0.001 _ {-0.497} ^ {+0.502}$ \\
    Scaled planetary radius $R_\mathrm{p}/R_{\star}$ &  $\mathcal{U}[0,0.5]$ & $0.048 \pm 0.002$  \\
    Impact parameter, $b$ &  $\mathcal{U}[0,1]$  & $0.285 _{-0.199} ^{+0.273} $ \\
    %

    \multicolumn{3}{l}{\emph{ \bf Other system parameters}} \\
    %
    Stellar density $\rho_\star$ ($\rho_{\odot}$) &  $\mathcal{N}[1.236, 0.209]$ & $1.159 _{-0.218} ^ {+0.215}$  \\
    Stellar density $\rho_\star$ (g cm$^{-3}$)  &   & $1.634 _{-0.308} ^ {+0.303}$  \\
    Limb darkening $q_1$ \tess\  & $\mathcal{U}[0,1]$ & $0.264 _{-0.180} ^ {+0.346}$ \\
    Limb darkening $q_2$ \tess\ & $\mathcal{U}[0,1]$ & $0.419 _ {-0.286} ^ {+0.351}$ \\
    \hline
    \multicolumn{3}{l}{\textbf{Derived parameters for \sname b}} \\
    Planet radius ($R_{\oplus}$)  & $\cdots$ & $ 4.242  _{-0.313} ^{+0.376} $ \\
    Scaled semi-major axis $a/R_\star$   & $\cdots$ & $ 11.732 _{-0.789} ^{+0.686} $ \\
    Semi-major axis $a$ (AU)  & $\cdots$ & $ 0.0498 \pm 0.0007 $ \\
    $e$  & $\cdots$ & $0.285_{-0.099}^{+0.133}$  \\
    $\omega_\star $ (deg)  &  $\cdots$ &  268 $\pm$ 46 \\
    Orbital inclination $i$ (deg)  & $\cdots$ & $88.6 \pm 1.0$ \\
    Transit duration (hours) & $\cdots$ & $2.761 _{-0.374} ^{+0.259}$ \\
     \multicolumn{3}{l}{\textbf{Derived parameters for \sname c}} \\
    Planet radius ($R_{\oplus}$)  & $\cdots$ & $4.793 _{-0.351} ^{+0.410}$ \\
    Scaled semi-major axis $a/R_\star$   & $\cdots$ & $20.728 _{-1.394} ^{+1.212}$ \\
    Semi-major axis $a$ (AU)  & $\cdots$ & $ 0.0880 \pm 0.0014 $ \\
    $e$  & $\cdots$ & $0.175_{-0.103}^{+0.139}$  \\
    $\omega_\star $ (deg)  &  $\cdots$ & $ 268 \pm 58$  \\
    Orbital inclination $i$ (deg)  & $\cdots$ & $89.2 \pm 0.6$ \\
    Transit duration (hours) & $\cdots$ & $3.723 _{-0.446} ^{+0.333}$ \\
     \hline
  \multicolumn{3}{c}{\textbf{RV fit}} \\
  \hline
  Parameter & Prior$^{(\mathrm{a})}$ & Value$^{(\mathrm{b})}$  \\
  \hline
    \multicolumn{3}{l}{\textbf{Parameters for \sname b}} \\
    Radial velocity semi-amplitude variation $K$ (m s$^{-1}$) &  $\mathcal{U}[0,100]$ & \textless 7 \\ 
    Planet mass ($M_{\oplus}$)  & $\cdots$ & $\textless$ 16    \\
    \multicolumn{3}{l}{\textbf{Parameters for \sname c}} \\
    Radial velocity semi-amplitude variation $K$ (m s$^{-1}$) &  $\mathcal{U}[0,100]$ & \textless 12 \\ 
    Planet mass ($M_{\oplus}$)  & $\cdots$ & $\textless$ 37   \\
    \multicolumn{3}{l}{\emph{ \bf Stellar activity GP model Parameters}} \\
    $h$  (\ms)  &  $\mathcal{U}[0.01, 1000]$ & $108.25_{-30.83}^{+48.06}$ \\
    $\lambda$  (days)  &  $\mathcal{U}[5, 2000]$ & $914.47 _{-653.08}^{+737.46}$ \\
    $\omega$    &  $\mathcal{N}[0.35, 0.035]$ & $0.36 \pm 0.03$ \\
    $\theta$  (days)  &  $\mathcal{N}[3.4, 0.5]$ & $3.37 _{-0.005}^{+0.006}$ \\
    Jitter term $\sigma_{\rm HARPS}$ (\ms) & $\mathcal{U}[0,100]$ & $65.400 _{-9.883} ^{+12.186}$  \\
  \hline
   \noalign{\smallskip}
  \end{tabular}
~\\
  \emph{Note} -- $^{(\mathrm{a})}$ $\mathcal{U}[a,b]$ refers to uniform priors between $a$ and $b$, $\mathcal{N}[a,b]$ to Gaussian priors with median $a$ and standard deviation $b$.\\  
  $^{(\mathrm{b})}$ Parameter estimates and corresponding uncertainties are defined as the median and the 16-th and 84-th percentile of the posterior distributions.\\
\end{table*}

\section{RV modeling}\label{sec:rv}
For the RV fit we employed the same package {\tt PyORBIT} as for the light curve fit. Given the small sample size, and being  the stellar activity the predominant signal in our dataset, a proper planet detection from RVs was not possible. For the same reason, we did not perform a joint fit with the light curve. On the other hand, we could infer an upper limit on the mass of both planets.

We tested five different models to fit the HARPS-N RV data: {\it i}) a circular two-planets with a Gaussian Process (GP) model (circ2p+GP) to fit the stellar activity with a quasi-periodic kernel {\it ii}) an eccentric two-planets with a Gaussian Process (ecc2p+GP); {\it iii}) same as {\it ii}) but with three planets (ecc3p+GP) to explore the possibility of an additional planetary companion; {\it iv}) same as {\it iii}) adding a linear trend to check on possible outer companion (ecc2p+GP+trend; see also Sec. \ref{sec:contamination}); {\it v}) a GP only model. The Gaussian process regression is performed through the package \texttt{george} \citep{Ambikasaranetal2015}; we employed the quasi-periodic kernel as defined by \cite{Grunblattetal2015}:

\begin{equation}
    h^2 \exp \biggl[-\frac{\sin^2{[\pi (t_i - t_j)/\theta]}}{2\omega^2} - \biggl(\frac{t_i - t_j}{\lambda}\biggr)^2\biggr] ,
\end{equation}

\noindent where $h$ represents the amplitude of the correlations, $\theta$ is the rotation period of the star, $\omega$ is the length scale of the periodic component, which is related to the size evolution of the active regions, and $\lambda$ represents the correlation decay timescale.

We ran the first four models performing a fit, which includes a Keplerian orbit for the planetary signal and independent jitter and offset terms. Using the orbital periods, the transit epochs obtained from the transit fit and the eccentricity (in the case of eccentric models) as Gaussian priors, and the stellar parameters obtained in Sec. \ref{sec:stellar}, we sampled the orbital period and the RV semi-amplitude in a linear space and followed the same parametrization as for the transit fit.
We ran the sampler for 100,000 steps, 128 walkers. The burn-in cut and thinning factor are the same as reported in Sec. \ref{sec:tess_analysis}. Also for the RV models, we computed the BIC and AICc criteria.
We reported the results in Table \ref{tab:rvmodelcomp}: we found that the GP only model is slightly preferred over the others. This is mainly due to the fact that the RVs cannot give a detection and are mostly dominated by the stellar activity. Moreover, the models with two planets are strongly preferred over the three planets case by both the BIC and the AICc, and in particular the circular model is favoured over the eccentric one. These results are a consequence of the lack of a significant detection and the strong penalty given to models with an higher number of free parameters by the BIC and AICc criteria. 

However, both circular and eccentric models return very similar parameters' values. We found a jitter term related to the stellar activity of $\sim$ 65\ms, and we assessed an upper limit to the RV semi-amplitudes, obtaining K$_b$ \textless\ 7 \ms\ and K$_c$ \textless\ 12 \ms\, corresponding to planetary masses of M$_{b}$ $\textless$ 16 \mearth\ for planet b, and M$_{c}$ $\textless$ 37 \mearth\ for planet c, at 1$\sigma$ confidence level. These parameters, together with the GP model parameters, are listed in Table \ref{tab:transit_params}. 

\subsection{Contamination from possible stellar companions and line of sight objects}
\label{sec:contamination}

In order to evaluate the possibility of an astrophysical false positive caused by an eclipsing binary blended with \sname\ in the \tess\ photometric aperture, we first considered the sources within
50 \arcsec\ in \gaia\ DR2. As seen in Fig. \ref{fig:vetting}, there is only one faint source, 2MASS J05063719-2014292 = TIC 146520534 at 23.4 \arcsec, which is however ruled out as responsible for the observed transit by the centroid test.


To estimate the chance for additional contaminants (either bound companions or field objects)
we adopted the \gaia\ DR2 detection limits derived by \citet{brandeker19}.
Considering targets with appropriate magnitude, the detection limits are of 
2.25 mag at 1.0 \arcsec\ and 9.0 mag at 4.0 \arcsec, corresponding to bound objects of mass 0.6 and $<0.1$ \msun, at 150 and 600 au, respectively.

We also used the {\tt TRILEGAL} model of the Galaxy \citep{2005A&A...436..895G} to simulate a population of stars along the line of sight: the number density of stars can be used to calculate the frequency of chance alignment given an aperture or radius of confusion. Using the Gaia contrast curve and the constraint from the transit depth of \sname\,c, we obtained a maximum radius of 2.4 arcsec. The {\tt TRILEGAL} simulation along the line of sight yields 821 bright enough stars per square degree. With the aforementioned radius, this yields an expected frequency of chance alignment is $\sim$0.1\%. Since the binary fraction is 33\% \citep{Raghavanetal2010} and the geometric transit probability for a 10.2 day orbit is $\sim$1.5\% (using the \sname's density and assuming a conservative maximum eclipse depth of 100\%), which represents the fraction of binaries with the same period that would be eclipsing as viewed from Earth, we obtained a probability of 0.0005\% that the signal is a background eclipsing binary (BEB). Repeating the same calculation for \sname\,b, we obtained a slightly higher probability of 0.0007\%. The probability of chance alignment with two different eclipsing binaries is the product of the two, which is extremely small, i.e. $\textless$\,10$^{-10}$. So the
likelihood is low that either signal is a BEB and very low that both are
BEBs, given a total number of TESS targets of about 200,000. However, this analysis is rather conservative, since it doesn't take into account the transit shape, which would further eliminate BEB scenarios with incompatible radius ratios.

From the available systemic RVs in Table \ref{tab:stellar}, small offsets are present 
between HARPS-N and \gaia\ DR2 and RAVE DR5. However, they are of marginal
significance (less than 2$\sigma$ in both cases, according to the nominal errorbars). Also, the HARPS-N RVs do not show significant trends within the timescale of our observations.
An upper limit on the RV slope of 0.73 m s$^{-1}$ d$^{-1}$ (1$\sigma$ confidence level) is obtained through a dedicated PyOrbit run including the presence of a linear trend.
The CCF of our HARPS-N spectra also appears without signatures
of additional components, although with the typical alterations of
young, spotted stars. 

In order to assess the potential presence of additional non-transiting companions, we computed the minimum-mass detection thresholds of our
HARPS-N RV time series. As previously done in \citet{Carleo2020a}, we followed the Bayesian approach from \citet{tuomi14} to compute the detectability function and detection thresholds: we applied this technique on the RV residuals after correcting for the correlation with the BIS, and included in the model the signals of the two planets as discussed in Sect. \ref{sec:rv}. Considering orbital periods between 0.5 and 200 d, we are sensitive to planets of minimum masses M$_{\rm P} \sin i > 0.40^{+0.11}_{-0.12}$ M$_{\rm J}$ , for $0.5 < P < 10$ d, and M$_{\rm P} \sin i > 0.96^{+0.46}_{-0.28}$ M$_{\rm J}$ for $10 < P < 200$ d. For longer periods, due to the short baseline of our RV observations, our sensitivity drops and we are not sensitive even to the larger sub-stellar companions. 

Moreover, when considering the proper motion of the star from the most relevant
astrometric catalogs such as \gaia\ DR2, \gaia\ DR1, Tycho2, PPMXL, SPM4.0, UCAC4, and UCAC5, no differences above 1.1 $\sigma$ are present,  no astrometric excess noise is reported in \gaia\ DR2, and the re-normalised unit weight error (RUWE) is 1.07, well below the threshold
of 1.4 indicating the need of additional parameters in the astrometric solution. All these results support the conclusion that there
are no close companions that might represent a source of astrophysical false positive,
or significantly dilute the observed transit depths, although the available detection limits 
do not allow to rule out all the potential stellar companions to \sname.




\subsection{False Positive probability}

We computed statistical false positive probabilities (FPPs) for \planetb\ and \planetc\ using the {\sc Python} package {\tt VESPA} \citep{2015ascl.soft03011M}. In brief, {\tt VESPA} computes the likelihoods of astrophysical false-positive scenarios involving eclipsing binaries by comparing the observed transit shape with simulated eclipsing populations based on the {\tt TRILEGAL} model of the Galaxy \citep{2005A&A...436..895G}. In particular, {\tt VESPA} explores three different false positive scenarios:  HEB (hierarchical eclipsing binary), EB (eclipsing binary) and BEB (background eclipsing binary -- physically unassociated with target star). For planet b, we find the FP probabilities $P_{HEB}$ = 0.06\%, $P_{EB}$ = 3.13\%  and $P_{BEB} \ll$ 10$^{-6}$. For planet c, all the FPPs are $<$10$^{-6}$. 

However, because {\tt VESPA} does not account for multiplicity, these FPPs are overestimated by at least an order of magnitude \citep{2012ApJ...750..112L, 2016ApJ...827...78S, 2018AJ....156...78L}. Additionally, since the RVs put a constraint on the masses that rules out EBs, the planet probability would increase to over 99\%. We thus consider both \planetb\ and \planetc\ to be statistically validated at the 99\% confidence level.

\section{Discussion} \label{sec:disc}

With both planets smaller than 5 \rearth\ radii, and mass upper limits of 16 and 37 \mearth, 
this system appears to be very appealing for further analyses. We performed a study to investigate the evolution of the planetary atmospheres (Sec. \ref{sec:atm_evol}), discussed the system architecture (Sec. \ref{sec:arch}) and the implications of the eccentric (Sec. \ref{sec:ecc}) and circular (Sec. \ref{sec:circ}) orbit cases.

\subsection{Atmospheric evolution simulations}\label{sec:atm_evol}
We studied the atmospheric evolution of both  planets  evaluating the mass loss percentage assuming circular orbits. The integrated stellar flux causing  photoevaporation differs in the case of an eccentric orbit by $(1-e^{2})^{-1/2}$, that is, by a factor of $\sim 4$\% in the case of $e=0.285$, implying that this approximation can be applied  in our case.
To estimate the atmospheric mass loss rate, we used the hydrodynamic-based approximation developed by~\cite{Kubyshkina2018},  including the evolution of the stellar XUV luminosity and of the mass and radius of each planet, but neglecting the atmospheric gravitational contraction.
In order to account for the X-ray stellar luminosity evolution, we used the prescriptions given in~\cite{Penz2008}, whereas for the extreme ultraviolet radiation the relation given in~\cite{SanzForcada2011} was used.  We underline that the above model for the XUV temporal evolution, provides  the evolution  of the total  X-ray luminosity distribution using a scaling law just for the mean value~\citep{Penz2008}. For young stars the observed spread in X-ray luminosities is associated with the spread of stellar rotation rates~\citep{Pizzolato2003}. The consequence of different rotation rates is that slow and fast rotators  remain in the saturation regime  for different time periods that go from about  10 Myr for slow rotators to about 300 Myr for fast rotators~\citep{Tu2015}, implying very different levels of high energy radiation at which planets are subjected.
We accounted for the evolution of the radius following~\cite{Johnstone2015}. First, we estimate the radius of the rocky core, $R_{\rm c}$, assuming that the density is equal to that of the Earth for both planets, then we obtain $R_{\rm c}= R_{\oplus} (M_{\rm pl}/M_\oplus)^{1/3}$. Assuming a hydrogen dominated atmosphere, using equation (3) of~\cite{Johnstone2015} and using the planetary radius given in Table~\ref{tab:transit_params}, we estimate the initial atmospheric mass fraction $f_{\rm at}=M_{\rm at}/M_{\rm pl}$; finally at each time step we update  $f_{\rm at}$ and the planetary mass in response to the  mass loss. Then using the new values for the mass and the atmospheric fraction we calculate the new radius. 

Since the values for the mass in Table~\ref{tab:transit_params} are  upper limits and given the high uncertainty on the age of the star, for each planet a set of simulations was performed for three different values of the planetary mass, $1\times M_{\rm ul}$, $\frac {1} {2} \times M_{\rm ul}$, $\frac {1} {3} \times M_{\rm ul}$ (where $M_{\rm ul}=16$ $M_\oplus$ for \sname\,b and $M_{\rm ul}=37 M_\oplus$ for \sname\,c), and for a stellar age of 30, 50 and 80 Myr. For each simulation the initial X-ray luminosity has been set at $L_X=10^{30.07}$ erg s$^{-1}$, i.e., the initial mass loss rate of the planets does not depend on the stellar age. The estimated initial atmospheric mass fractions for the three planetary masses in the case of \sname\,b are 0.27, 0.19, 0.14, respectively;  in the case of  \sname\,c are 0.49, 0.5, 0.4, respectively.

For \sname\,b the calculated current mass-loss rates are $ 1.31\times 10^{13}$ g s$^{-1}$, $2.05\times 10^{14}$ g s$^{-1}$, $1.03\times 10^{15}$ g s$^{-1}$, for the three masses  $1\times M_{\rm ul}$, $\frac {1} {2} \times M_{\rm ul}$, $\frac {1} {3} \times M_{\rm ul}$ respectively. In the case of \sname\, c, for the masses $\frac {1} {3} \times M_{\rm ul}$ and $\frac {1} {2} \times M_{\rm ul}$, we derived current mass-loss rates of  $4.8\times 10^{12}$ g s$^{-1}$ and $1.2\times 10^{12}$ g s$^{-1}$, respectively, while for $1 \times M_{\rm ul}$ the planet is stable against hydrodynamic evaporation and the mass loss rate is negligible because the atmospheric losses are limited to Jeans escape. 

\begin{figure}
\centering
\includegraphics[width=.55\textwidth]{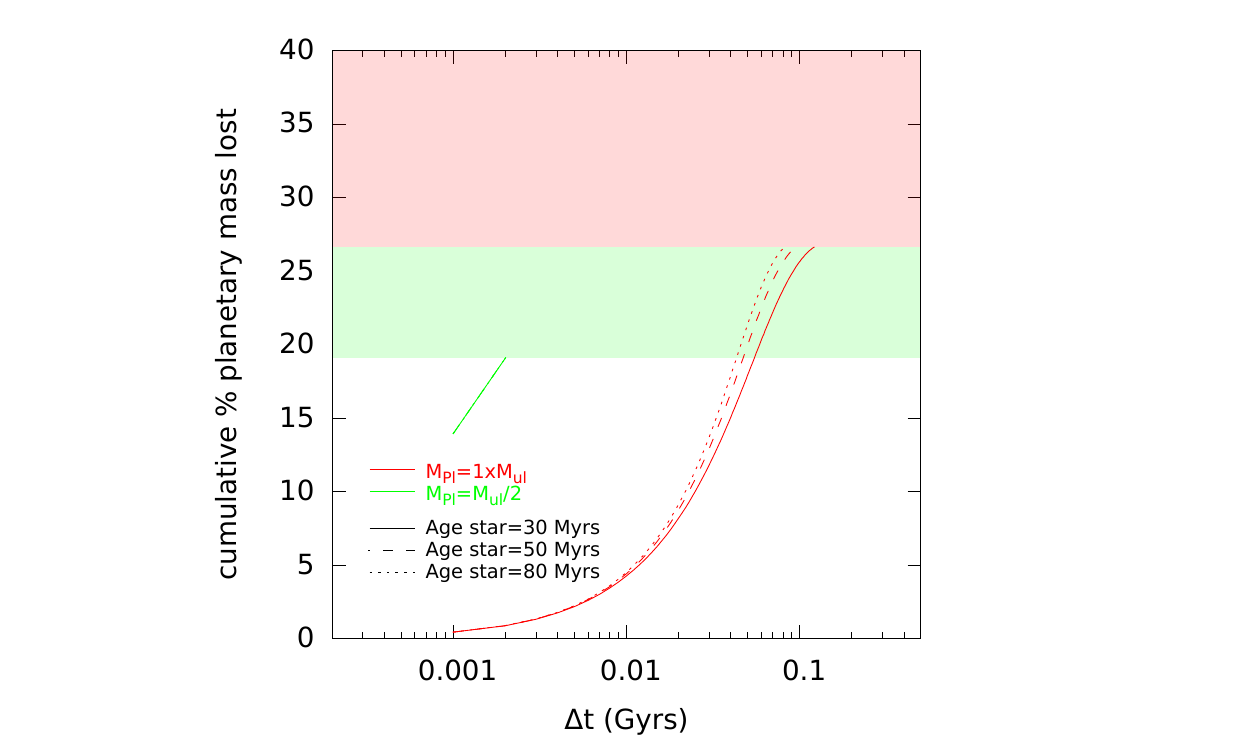} 
\includegraphics[width=.55\textwidth]{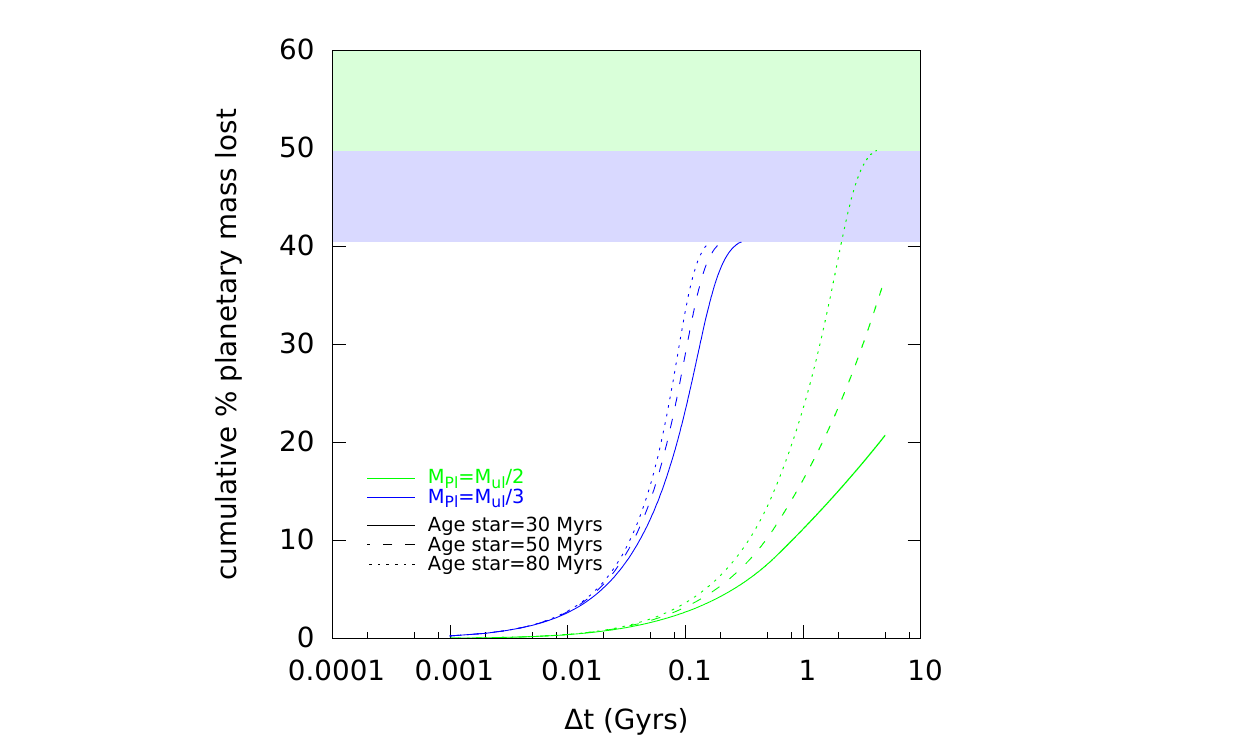} \\
\caption{Cumulative percentage mass loss as function of $\Delta t = t-T_{\rm age}$. 
Red, green, and blue lines refer to planetary masses of 
$1\times M_{\rm ul}$, $\frac {1} {2} \times M_{\rm ul}$, and 
$\frac {1} {3} \times M_{\rm ul}$, respectively.
Stellar ages of 30, 50, and 80 Myr are shown in solid, dashed, and dotted lines, respectively. The shaded areas represent the threshold of mass loss: above this limit, the planet cannot lose any further mass. 
\textit{Upper panel}: \sname\,b. \textit{Bottom panel}: \sname\,c.}
\label{fig:ev_cumu}
\end{figure}

 
 Figure~\ref{fig:ev_cumu} shows the cumulative mass loss percentage as a function of $\Delta t = t-T_{\rm age}$ (where $t$ is the time and $T_{\rm age}$ the stellar age). In general, we found that, as expected, in the case of high  $f_{\rm at}$ (high mass) the planet takes a longer time to lose its atmosphere; on the other hand, for a given mass,  older stellar ages translate to shorter times  to lose its envelope. This is basically due to the fact that the X-rays luminosity decays slower in time at older ages. The time taken to lose entirely the atmosphere goes from few Myr in the case of the lowest masses to Gyr in the case of highest masses. In particular, in the case of $\frac {1} {3} \times M_{\rm ul}$ \sname\,b loses its atmosphere in less than 1 Myr, while in the case $\frac {1} {2} \times M_{\rm ul}$, the planet evaporates in about 2 Myr. 
 
 \sname\,c \textit{i)} in the case of $\frac {1} {3} \times M_{\rm ul}$,  evaporates completely its atmosphere in around 200-300 Myr, depending on the stellar age; \textit{ii)}  in the case of $\frac {1} {2} \times M_{\rm ul}$ it loses its envelope only for the stellar age  of  80 Myr in around  4.2 Gyr;
 while for the stellar ages of 30 and 50 Myr it loses only fractions of its atmosphere in 5 Gyr; \textit{iii)} in the case of $1 \times M_{\rm ul}$ its atmosphere is hydrodynamically stable and the planet can lose only negligible amounts of its atmosphere  through Jeans escape (hydrostatic evaporation). 
 
When the planets entirely lose their atmospheres, the final planetary radius value is given by the core radius value, which depends on the initial value of the planetary mass. On the other hand, in the cases in which the planets lose only fractions of their envelope, the final radius value depends on equation (3) of~\cite{Johnstone2015}. Generally, the radius distribution of close-in super Earths and sub-Neptunes follows a bi-modal distribution (for details see~\citealt{Fulton2017} or~\citealt{Modirrousta2020}). As expected for this kind of planets, in the cases of  $\frac {1} {2}$, $\frac {1} {3}$ $\times M_{\rm ul}$ \ and for all stellar ages, the radius of \sname\,b,  which initially lies on the right peak of the distribution, crosses the radius gap and ends its temporal evolution at the base of the left peak,  which is likely populated by bare core planets.

\subsection{System architecture}\label{sec:arch}

We briefly discuss here the planetary system around \sname\ compared to other systems.
It is interesting to understand whether young planetary systems have distinct features with respect to the mature ones that might shed light on their evolution. Of course, analysis based on one individual system might be biased. A well-defined sample of young planetary systems is mandatory for a statistical evaluation.

\citet{weiss2018} studied 909 planets in 355 multi-planet systems observed by \kepler, finding interesting and definite correlations among the characteristics of the planets. In particular, they found that planets in a multi-planet system present correlated masses or radii and in 65\% of cases the outer planets are larger than the inner planets. \sname\,b and c, with  radii of 4.3 and 4.8 \rearth\ respectively, follow the same picture. 
As discussed in Sec. \ref{sec:atm_evol}, the planets of \sname, especially the planet b, are expected to suffer from significant photoevaporation which changes their radii  with time. Considering the evaporation model introduced in Sect.~\ref{sec:atm_evol}, the outer planet is expected to have a larger radius than the inner planet at all evolutionary stages. The larger relative shrinking of the radius of \sname\,b would change significantly the ratio of the radii along the system evolution, but this should
remain within the distribution of systems studied by \citealt{weiss2018}, considering the difference in planet temperature.

We also examined the planetary separation in terms of mutual Hill radii ($R_{\rm H}$), in order to understand how far the two planets formed from each other. We estimated the value of the planetary spacing as in \citealt{weiss2018}, and we found that \sname\ planets have a separation of $\sim$ 17 $R_{\rm H}$, in agreement with \citealt{weiss2018} results which show that 93\% of planet pairs are at least 10 mutual Hill radii apart\footnote{To be homogeneous with the analysis of \cite{weiss2018} 
this estimate is derived for circular orbits and estimating the planetary masses from the empirical mass-radius relationships reported in \cite{weiss2018}. }.

Moreover, we investigated the possible mean-motion resonances between the two planets. According to \citet{fabrycky2014}, most pairs of planets are not in mean-motion resonances. The period ratio between \sname\,b and c is 2.349. This is close to the 7:3 ratio (2.333),
corresponding to a minor peak in \citet{fabrycky2014} distribution.

\subsection{Implications of eccentric orbits}\label{sec:ecc}
We here discuss some implications that possible eccentric orbits for both planets can have on this system and our understanding of its history. 
\sname\,b and c join the small group of Neptune-type planets with orbital periods of a few days around late-type stars.  Those planets often present non-negligible orbital eccentricities, especially the subgroup with orbital periods shorter than $\sim$10~days for which the mean value is  $\sim$0.15-0.20.
\begin{figure}[t]
   \centering
\includegraphics[width=1.0\linewidth, trim = 0 13cm 0 4cm, clip]{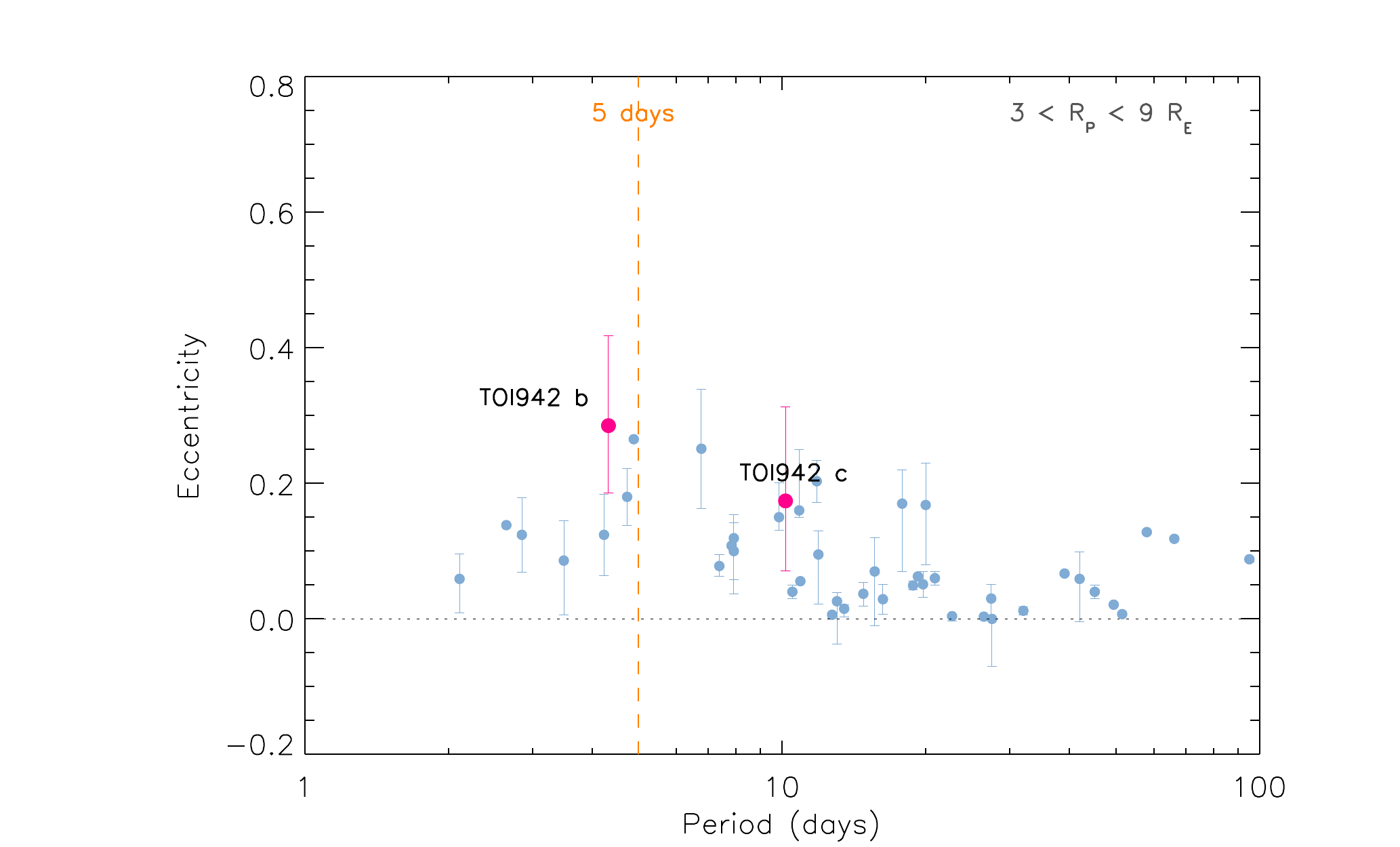}
\caption{Distribution of eccentricities as a function of orbital period as shown in \cite{correia2020}, for Neptune size planets. The eccentricities have uncertainties smaller than 0.1. \sname\,b and c are overplotted in pink.}
\label{fig:toi942correia}
\end{figure}
A typical non-zero eccentricity for Neptunes in close orbits was also pointed out by \cite{correia2020}.
The behaviour of Neptune-radius planets in the period-eccentricity diagram is different both with
respect to giant planets, which show a clear increase of eccentricity at period longer than 5 days
and, to smaller planets (R $\textless$ 3 \rearth), which have typically low eccentricity.

In order to understand the position of \sname\,b and c in the same diagram, we reproduced the middle panel of Fig. 1 in \cite{correia2020}. We used Exo-MerCat \citep{alei2020}, a Python software that merges all the information from the four exoplanets' catalogues, NASA Exoplanet Archive\footnote{\url{https://exoplanetarchive.ipac.caltech.edu/}} \citep{Akesonetal2013}, Exoplanet Orbit Database\footnote{\url{http://exoplanets.org/}} \citep{Wrightetal2011}, Exoplanet Encyclopedia\footnote{\url{http://exoplanets.eu/}} \citep{Schneideretal2011}, and Open Exoplanet Catalogue\footnote{\url{http://www.openexoplanetcatalog.com/}} \citep{Rein2012}, in order to have a unique, uniform and standardized catalogue. The Exo-MerCat catalogue is publicly available as a VO resource\footnote{IVOID \url{ivo://ia2.inaf.it/catalogues/exomercat} served by the TAP service \url{ivo://ia2.inaf.it/tap/projects}} and it is updated weekly. We selected planets with orbital period 1 $\textless$ P$_{\rm orb}$ $\textless$ 100 days, planetary radius 3 $\textless$ R$_{\rm P}$ $\textless$ 9 R$_{\oplus}$ and eccentricity with uncertainties smaller than 0.1. Then we considered the eccentricities obtained for \sname\,b and c from the transit fit (ecc2p) and added them to the sample (pink dots, Fig. \ref{fig:toi942correia}). They follow the distribution within their uncertainties, although it is worth to stress that our resulting eccentricities cannot be well constrained from our data and while the circular model is preferred over the eccentric one, the latter leads to a consistent stellar density. Consequently, we cannot give any definitive conclusion on this aspect. An extensive RV survey would be needed to have a complete determination of the planet characteristics.


Assuming a modified tidal quality factor $1.6 \times 10^{5} \la
Q^{\prime}_{\rm p} \la 5.6 \times 10^{5}$ as suggested by the
evolutionary scenario of the orbits of the main satellites of Uranus
\citep{TittemoreWisdom90,Ogilvie14}, we estimate an e-folding decay
timescale for the eccentricity \footnote{The e-folding decay timescale
is defined as $e/(de/dt)$, where $e$ is the eccentricity and $t$ the
time, and is calculated for the present values of the system
parameters.} of planet b ranging from 0.8 to 2.7 Gyr, while for planet c
it ranges from 62 to 225 Gyr owing to the rapid decay of the tidal
effects with the increase of the semi-major axis of the planetary orbit.
A consequence of tidal dissipation is the internal heating of planet b
that provides a surface flux of about 575 W~m$^{-2}$ for
$Q^{\prime}_{\rm p}=1.6 \times10^{5}$ and scales in inverse proportion
to the value of the tidal quality factor. It is much larger than in the
case of Jupiter that shows a flux of only 5.4~W~m$^{-2}$
\citep{Guillotetal05}. In the case of planet c, the tidally induced flux
ranges from $\sim 0.8$ to $\sim 3$~W~m$^{-2}$ for the adopted range of
$Q^{\prime}_{\rm p}$, comparable with the value of the internal heat
flux in Jupiter. 

\begin{figure}
\includegraphics[width=01.0\linewidth]{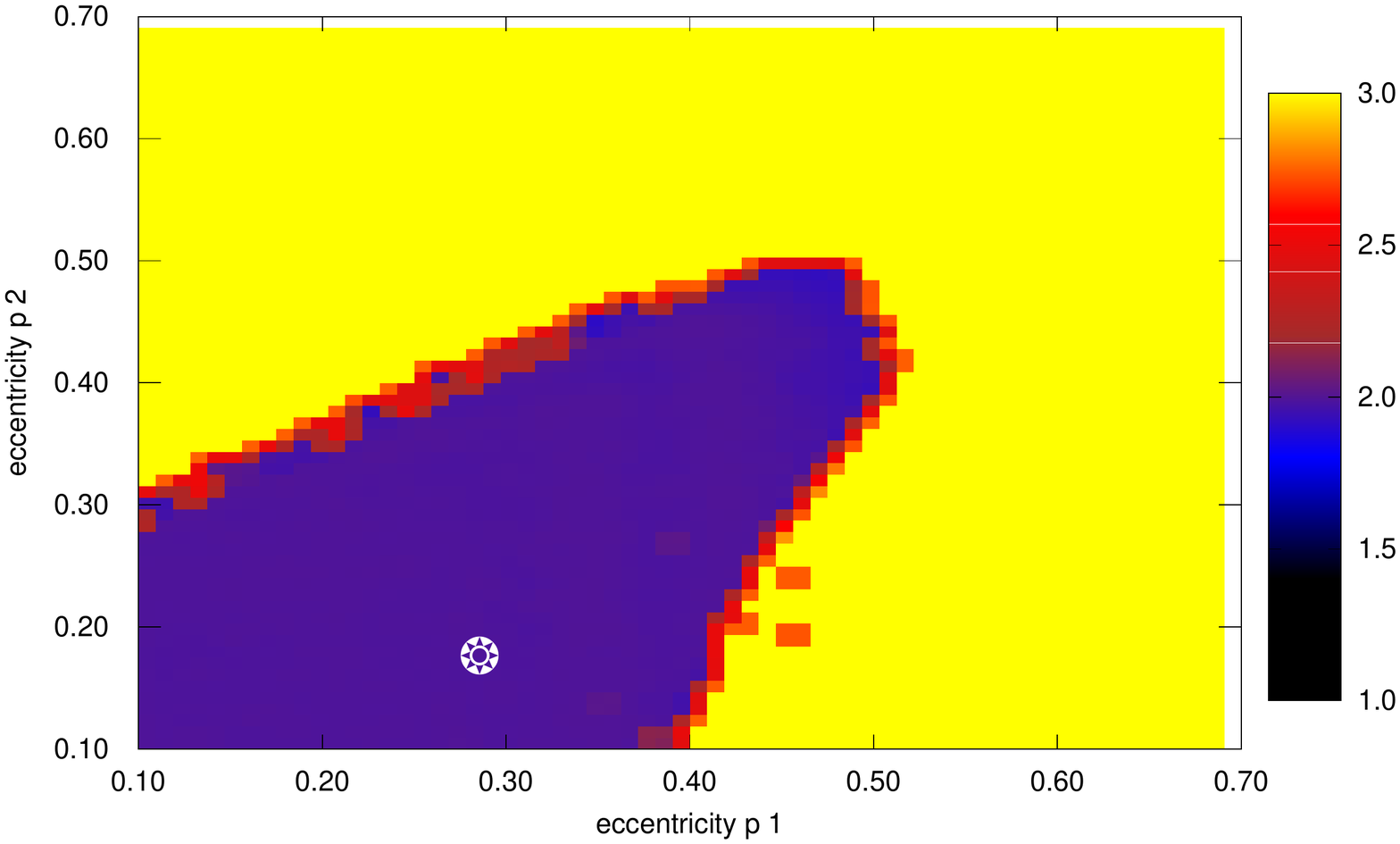}
\caption{MEGNO indicator computed for different initial values of the orbital eccentricity of the two planets. The color bar indicates the stability scale: the values of the MEGNO for the stable area is close to 2. Larger values of the MEGNO indicator (the yellow region) point to chaotic evolution. The star symbol shows the nominal eccentricities values.}
\label{fig:megno}
\end{figure}

To gain insight into the implications of the eccentric model, we have first verified the values of eccentricities for which \sname's system could be stable through the Mean Exponential Growth factor of Nearby Orbits MEGNO \citep{cincotta2000, god2008}.
MEGNO  is  closely  related  to  the  maximum  Lyapunov exponent, providing an alternative determination of it. In case of regular or quasi--periodic motion the MEGNO indicator is $\approx$2 while for chaotic motion it increases with time. To test the stability around the nominal orbit, we have regularly sampled the initial eccentricities of the two planets in between 0.1--0.7 and computed the MEGNO for each orbit.
In the numerical integrations, spanning 50 Kyr, the initial semi--major axis and periastron argument of each planet are set to the nominal values and the mutual inclination is equal to 0 since both planets  
transit the star.\\ 
\indent To test the most difficult conditions for the dynamical stability of the system, we adopted the highest values for each planet mass, i.e. $m_1 = 16$ $M_{\oplus}$ and $m_2= 37$ $M_{\oplus}$ (see Table 2).  The results are shown as a  stability map in Fig.\ref{fig:megno}. The stable area (blue region in the plot), where the values of MEGNO are close to 2, extends up to about 0.5 in eccentricity for both planets and the nominal solution is well within the stable region suggesting that the high eccentricities derived from the system are not critical for its long term stability.

We then investigated \sname's dynamical history by means of its normalized angular momentum deficit (NAMD, \citealt{chambers2001,turrini2020}), an architecture-agnostic measure of the dynamical excitation of a planetary system. The NAMD allows for comparing the dynamical excitation of planetary systems with diverse architectures and for gaining insight on the differences in their dynamical histories \citep{turrini2020}. We took advantage of this property to compare the dynamical excitation of \sname\ with that of two template systems \citep{turrini2020}: Trappist-1 \citep{gillon2017,grimm2018} and the Solar System. Trappist-1's dynamical history was shown to be characterized by stable and orderly evolution shaped by orbital resonances and tidal forces \citep{tamayo2017,papaloizou2018}. The Solar System, on the other hand, lies at the boundary between orderly and chaotic evolution, with signs of chaos and long-term instability in its current architecture and possible past phases of dynamical instability \citep[e.g.][]{laskar2017,nesvorny2018}.

To compute the average NAMD value of \sname\ taking into account the uncertainty in its physical and orbital parameters, we followed the Monte Carlo approach described by \citet{laskar2017} and \citet{turrini2020}. We performed $10^{4}$ Monte Carlo extractions of the physical and orbital parameters of \sname's planets and used them to compute the NAMD value of the resulting $10^{4}$ simulated systems. For all parameters, we assumed standard deviations equal to half the confidence intervals of the respective quantities \citep{laskar2017}. Following \citet{zinzi2017} and \citet{turrini2020}, we adopted as \sname's reference plane the orbital plane of the largest planet, \sname\,c, and converted the inclinations to relative inclinations with respect to this plane. 
As we possess only upper limits for the planetary masses we assumed the two planets to have similar densities (by analogy with Uranus and Neptune in the Solar System) and used their volumes in  computing the NAMD \citep[see also][]{he2020}. For the orbital eccentricities, we considered the posterior distributions for the planetary eccentricities shown in Fig. \ref{fig:ecc} truncated between zero and 0.5 to account for the results of the stability study with MEGNO.

\begin{figure}[t]
   \centering
\includegraphics[width=\hsize]{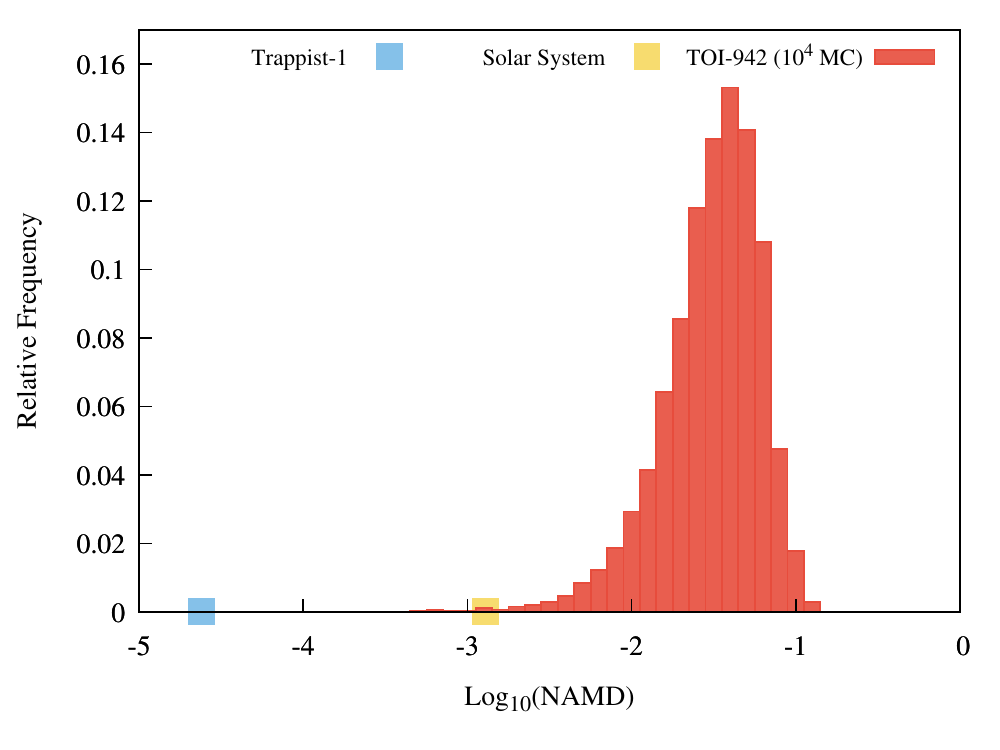}  
\caption{NAMD lognormal distribution of the $10^{4}$ Monte Carlo samples of \sname\ computed varying the orbital and physical parameters of the two planets within their confidence intervals (see main text for details). Also shown are the NAMD values of the Solar System (orange square) and Trappist-1 (blue square) for comparison. The horizontal positions of the Solar System and of Trappist-1 are arbitrary.}
\label{fig:toi942-namd}
\end{figure}

Fig. \ref{fig:toi942-namd} shows the Monte Carlo lognormal distribution of \sname's NAMD. The mean NAMD value is $3\times10^{-2}$ with the 3$\sigma$ confidence interval extending from $4\times10^{-3}$ to $0.3$. Fig. \ref{fig:toi942-namd} also shows the NAMD values of Trappist-1 ($2.4\pm0.4\times10^{-5}$) and of the Solar System ($1.3\times10^{-3}$). We refer readers to \citet{turrini2020} for details on their computation. The NAMD values of both Trappist-1 and the Solar System fall well below \sname's confidence interval indicating that in the eccentric model, even if \sname\ is currently dynamically stable, its dynamical history was more violent and chaotic than those of the other two systems. Using the full range of eccentricities and letting the planetary masses vary between $M_{ul}$ and $1/3 M_{ul}$ (see Sect \ref{sec:atm_evol}) produces an analogous result, albeit with larger values for the mean NAMD and the upper boundary of the 3$\sigma$ confidence interval. It is interesting to note that the period ratio close to 7:3 of \sname~b and c supports the picture depicted by \sname's NAMD. The dynamical characterization of the 7:3 resonance in the asteroid belt \citep{gladman1997} shows how its timescale of ejection is of the order of a few tens of Myr, i.e. shorter than \sname's age. If the two planets were originally trapped in a resonant condition \citep[e.g.][and references therein]{xu2017}, the eccentricity jump associated with their exit from it could be the violent dynamical event recorded by \sname's NAMD value.

\subsection{Implications of circular orbits}\label{sec:circ}
If future investigations will establish that \sname\,b and c have circular orbits, they could have migrated through a protoplanetary disc via a type I migration \citep[e.g.,][]{Nelson18}. The possibility that they formed via high-eccentricity migration is at variance with the very long circularization timescales, especially for planet c. Therefore, the possibility of explaining the  eccentric orbits observed in similar systems as the residual of their formation through a high-eccentricity migration followed by sizeable evaporation of their atmospheres \citep{correia2020} does not appear applicable to our system in the case of circular orbits because the eccentricity of planet c did not change appreciably during its lifetime.

Testing the NAMD value of the planetary system in the circular case, i.e. accounting only for the dynamical excitation due to the relative inclination of the planets, returns a value only slightly higher than the one of Trappist-1, suggesting a orderly evolution to the current architecture and further excluding the possibility of high-eccentricity migration.

\section{Conclusions and Future Perspectives}\label{sec:concl}
In this paper we presented the validation of the Neptune-type planet and the discovery of a second Neptune transiting the young star \sname\ (TYC 5909-319-1, TIC 146520535), observed by \tess\ in Sector 5, with periods of 4 and 10 days, respectively. Thanks to \tess, REM, SuperWASP photometry and HARPS-N spectroscopy we constrained most of the main stellar and planetary parameters. \sname\ is a young and relatively active star with an age of $50_{-20}^{+30}$ Myr and an activity index of \logrhk\ = -4.17\,$\pm$\,0.01. \sname\,b and c are Neptune-type planets with a radius of 4.3 and 4.8 \rearth, and a mass upper limit of 16 and 37 \mearth, respectively. While the RV data do not present planet detections and are only used to infer an upper limit on the planetary masses because of the high stellar-activity jitter, the \textit{TESS} light curves coupled with complementary spectroscopic,
astrometric  and imaging datasets allow for system validation. Although the circular transit model is favoured over the eccentric one, it brings to a stellar density value which is inconsistent with the stellar parameters obtained from the spectroscopy. This inconsistency disappears when the eccentricity is included in the model. In this case, we found a slightly non-zero eccentricity for the planet b. However, we stress the fact that the eccentricity distribution for each planet is the outcome of geometrical constraints (i.e., transit duration, impact parameter and stellar density), since the poor sampling of the ingress/egress and the lack of a secondary eclipse do not allow for a precise determination of this parameter.
Further RV observations are definitely important to better characterize the planetary masses and eccentricities, which will allow studying the dynamical and evolution history of this system.

Our evaluations on planetary mass loss suggest that this system is very interesting for future follow-up observations and atmospheric characterization. These kinds of systems with more than one planet plays a very crucial role in understanding the physics behind the planetary formation process, and when all planets transit the star we have a great opportunity to obtain a comprehensive characterization of the system. We plan to keep monitoring this star in our GAPS program. Moreover, being \sname\ part of CHEOPS sample, we will soon have very high-precision observations that will allow us to better refine the planetary radii, as well as investigate the transit timing variations (TTV) to explore the possibility of additional companions. 
The measurement of the Rossiter-McLaughlin effect would allow checking the relative
orientation of the planetary orbits and of the stellar spin. This is particularly relevant considering the young age and the possible dynamical history of the system.

\begin{acknowledgements}
The authors became aware of a parallel effort on the characterization of \sname\ by Zhou et al. [proper citation post-publication] in the late stages of the manuscript preparations. The submissions are coordinated, and no analyses or results were shared prior to submission.

This work has made use of data from the European Space Agency (ESA) mission {\it Gaia} (\url{https://www.cosmos.esa.int/gaia}), processed by the {\it Gaia} Data Processing and Analysis Consortium (DPAC,
\url{https://www.cosmos.esa.int/web/gaia/dpac/consortium}). Funding for the DPAC has been provided by national institutions, in particular the institutions participating in the {\it Gaia} Multilateral Agreement. We acknowledge financial support from the ASI-INAF agreements n. 2018-16-HH.0 and 2018-22-HH.0, as well as the INAF Main Stream project \textit{ARIEL and the astrochemical link between circumstellar discs and planets} (CUP: C54I19000700005). This paper includes data collected by the \tess\ mission. Funding for the \tess\ mission is provided by the NASA Explorer Program.
\end{acknowledgements}

\begin{appendix}
\section{Comoving objects}
\label{sec:comoving}

We looked for comoving objects around \sname\ in order to have additional constraints on stellar age and better
characterize the environment of the planet host. We queried \gaia\ DR2 catalog within 2 deg from the target for
objects with parallax difference smaller than 1 mas and proper motion difference smaller than 2 $mas yr^{-1}$.
Seven objects match these criteria (Table \ref{tab:comoving}).
Star \#7 has moderately blue colors from \gaia\ and Pan-STARRS \citep{panstarrs1} and its position on color-magnitude diagram is not compatible with a main sequence or pre-main sequence object; it lies slightly below the white-dwarfs sequence for the nominal parameters, but the astrometric parameters are highly uncertain. There is also a significantly brighter object at about 8" (2MASS J05104749-1913475),
that may bias photometric measurements. 

The position on CMD of Fig.~\ref{fig:cmd} shows that stars \#3 and \#4 (which are actually forming a wide binary with a projected separation of 13.3\arcsec $\simeq$ 2100 au) are well above the standard main sequence and close to the empirical  locus of Tuc-Hor, Columba, and Carina associations, suggesting they are young.
Their age appears fully compatible with our estimate for \sname.
Stars \#1, \#2, \#5, and \#6 are instead close to the main sequence and could be older interlopers.
The low absolute proper motion of \sname\ might allow significant contamination by unrelated objects.

None of the targets has RV measurements from \gaia\ or other sources or signatures of being young, as X-ray emission or UV excess due to chromospheric activity from GALEX. Only for stars \#3 and \#4  there are indications of photometric variability: they are classified as RR Lyr candidates in \citet{stringer2019}. This classification is 
clearly not compatible with the position on CMD from \gaia\ but can be the signature of short-period variability, considering the sparseness of their photometric measurements. 
We derived the photometric timeseries for all comoving candidates but star \#7 because of its faintness. Significant variability is detected for stars \#3 and \#4 (which are blended in the 
\tess\ data), with a possible periodicity of 0.47\,d. This period would fit nicely the
color-\prot\ sequence of the Pleiades \citep{rebull2016}, especially if the observed period
belongs to the brighter component (star \#4). We conclude from the position of CMD and 
photometric variability that the wide binary system composed by 2MASS J05064475-1835567  and 2MASS J05064509-1836091  is likely coeval and comoving with \sname.


\begin{sidewaystable}
\begin{footnotesize}
\addtolength{\tabcolsep}{-2.0pt}
\centering
\caption{Comoving objects from \gaia.}
\label{tab:comoving}
\begin{tabular}{lrrrrrrr}
\hline
Star ID         &   1 & 2 & 3 & 4 & 5 & 6 & 7  \\
\gaia DR2 ID    & 2962780178650659456 & 2976534901614084864 & 2976464704668708352 & 2976464498510278400  & 2974869278935546624  & 2974978607326043008  & 2975436248977764864 \\
2MASS ID        & J05101639-2108089   & J05052538-1816169   & J05064475-1835567   & J05064509-1836091    & J05052499-2032083    & J05020671-2030195     &   -- \\
TIC ID          & 146595452           & 146515819           & 146523357           & 146523356            &           146516626            & 146438186 	     & 671234760\\
\hline
\noalign{\smallskip}
separation (")  & 4454 & 7176 & 5928 & 5916 & 1443 & 3898 & 5093 \\
separation (pc) &  3.3 &  5.3    &  4.4    &    4.4  &  1.1    &   2.9   &  3.8    \\
$\pi$ (mas) 	                &  $6.0689\pm0.3223$ &  $6.7287\pm0.0577$ & $6.4258\pm0.0857$ &  $6.1935\pm0.1665$   &        $5.5275\pm0.0500$  &   $5.8237\pm0.0784$  &    $7.3210\pm1.8081$ \\
$\mu_\alpha$ (mas\,yr$^{-1}$) 	&  $15.479\pm0.396$  &	$15.001\pm0.064$  & $16.634\pm0.105$  &   $17.251\pm0.202$  &        $17.596\pm0.068$ &    $18.073\pm0.103$     &    $12.562\pm1.651$ \\
$\mu_\delta$ (mas\,yr$^{-1}$) 	&  $-4.629\pm0.464$  &  $-4.222\pm0.084$  & $-3.891\pm0.115$  &  $-3.708\pm0.215$    &       	$-1.977\pm0.075$  &    $-4.908\pm0.118$   &  	$-4.710\pm2.705$ \\
$\Delta \pi$ (mas) 	& -0.46 & +0.20 & -0.10 & -0.33 & -1.00 & -0.70 & +0.80 \\          
$\Delta \mu_\alpha$ (mas\,yr$^{-1}$) 	&  +0.10 & -0.38 & 1.25 & 1.87 & 2.21 & 2.69 & -2.82 \\
$\Delta \mu_\delta$ (mas\,yr$^{-1}$) 	&  -0.65 & -0.25 & 0.09 & 0.27 & 2.00 & -0.93 & -0.73 \\
\hline
\multicolumn{3}{l}{\it Optical and near-infrared photometry} \\  
\noalign{\smallskip}
$\tess$          & $17.792\pm0.019$  & $14.589\pm0.007$ &  $15.615\pm0.008$  & $14.905\pm0.008$ & $14.834\pm0.007$  & $15.731\pm0.008$ &  $20.696\pm0.008$   \\
\noalign{\smallskip}
$G$				 &  19.274 & 15.7317  & 17.0672 & 16.3535  & 16.0082   & 16.9814  & 20.9030 \\
$BP-RP$          &  3.2041 &  2.3309  &  3.3083 &  3.2462  &  2.4171   &  2.6361  & 0.4002 \\
\noalign{\smallskip}
$J$ 			&  $15.785\pm0.074$ & $13.181\pm0.026$ & $13.696\pm0.029$ & $13.061\pm0.029$ & $13.405\pm0.035$ & $14.158 	\pm0.036$ & -- \\
$H$				&  $15.426\pm0.127$ & $12.602\pm0.023$ & $13.180\pm0.028$ & $12.547\pm0.028$ & $12.737\pm0.035$ & $13.653 	\pm0.032$ & -- \\
$Ks$			&  $14.882\pm0.136$ & $12.361\pm0.026$ & $12.953\pm0.033$ &	$12.232\pm0.031$ & $12.526\pm0.034$ & $13.417 	\pm0.044$ & -- \\
\noalign{\smallskip}

%
\mstar\ ($\mathrm{M_\odot}$) & $^{+}_{-}$ & \\
\hline
\end{tabular}
\end{footnotesize}
\end{sidewaystable}

\clearpage

\onecolumn
\begin{longtable}{l|ccrc|rc}
\caption{\label{tab:YO38RVharpsn} Time series of \sname\ from HARPS-N data. We list radial velocities (RV), \logrhk, and their related uncertainties from DRS calculated through Yabi, and RVs together with uncertainties from TERRA pipeline.}\\
\hline
\noalign{\smallskip}
            &  \multicolumn{4}{c}{DRS}      & \multicolumn{2}{|c}{TERRA} \\ 
\hline
\noalign{\smallskip}
\multirow{2}{*}{JD-2450000}  &      RV  & $\sigma_{\rm RV}$  &   $\rm log\,R^{\prime}_\mathrm{HK}$ & $\sigma_{\rm log\,R^{\prime}_\mathrm{HK}}$ &    RV &    $\sigma_{\rm RV}$ \\ 
             & (\kms) &    (\kms)   & & & (\kms)   &  (\kms)  \\
\hline             
\noalign{\smallskip}
 8746.7492210  &     25.2430  &  0.0208  &  -4.1717 & 	0.0127 &   -0.0073 &       0.0129 \\
 8747.7425196  &     25.0373  &  0.0137  &  -4.1427 & 	0.0064 &  -0.1385 &       0.0111 \\
 8807.6666239  &     25.0627  &  0.0117  &  -4.1766 & 	0.0050 &  -0.1566 &       0.0094 \\
 8819.5844682  &     25.4891  &  0.0127  &  -4.1709 & 	0.0051 &  0.1705   &      0.0089 \\
 8831.5490394  &     25.1783  &  0.0199  &  -4.1896 & 	0.0110 &  -0.1004  &      0.0143 \\
 8838.6041231  &     25.1320  &  0.01559 &  -4.1948 & 	0.0081 &  -0.1102  &      0.0114 \\
 8841.5451455  &     25.2028  &  0.01742 &  -4.1310 & 	0.0080 &  -0.0707  &       0.0135 \\
 8845.4915917  &     25.1314  &  0.01861 &  -4.1710 & 	0.0098 &  -0.1125 &       0.0124 \\
 8846.4274669  &     25.5392  &  0.01303 &  -4.1783 & 	0.0061 &  0.2122  &       0.0104  \\
 8850.5058415  &     25.3127 &   0.01555 &  -4.1484 & 	0.0072 &  0.0469  &        0.0110 \\
 8851.5186877  &     25.1500 &   0.01529 &  -4.1190 & 	0.0065 &  -0.0609  &       0.0099 \\
 8853.4728514  &     25.5032 &   0.01277 &  -4.1690 & 	0.0053 &  0.1747  &        0.0095 \\
 8858.5110715  &     25.1471 &   0.02323 &  -4.2072 & 	0.0141 &  -0.1549 &       0.0157 \\
 8859.4653003  &     25.1183 &   0.02380 &  -4.2127 & 	0.0142 &  -0.1403 &        0.0168  \\
 8860.5200444  &     25.4354 &   0.01186 &  -4.1723 & 	0.0051 &  0.1457  &        0.0081 \\
 8861.4840393  &     25.2115 &   0.01309 &  -4.1420 & 	0.0056 &  -0.0402 &       0.0114  \\
 8884.4039209  &     25.4247 &   0.02547 &  -4.2055 & 	0.0161 &  0.0943  &        0.0168 \\
 8885.3985895  &     25.0756 &   0.01549 &  -4.1429 & 	0.0072 &  -0.1622   &     0.0104 \\
 8886.3772865  &     25.3471 &   0.01315 &  -4.2067 & 	0.0058 &  0.0809  &        0.0098 \\
 8887.3756575  &     25.2407 &   0.00893 &  -4.1736 & 	0.0035 &  0.0000   &         0.0067 \\
 8888.3795138  &     25.4760 &   0.01798 &  -4.1705 & 	0.0089 &  0.1111  &       0.0127  \\
 8889.4205776  &     25.2058 &   0.01206 &  -4.1895 & 	0.0052 &  -0.0445  &       0.0081  \\
 8891.3381780  &     25.4161 &   0.01222 &  -4.1790 & 	0.0055 &  0.1326 &        0.0085 \\
 8906.3778797  &     25.1067 &   0.01862 &  -4.1758 & 	0.0100 &  -0.1121 &       0.0125   \\
 8908.3837860  &     25.3920 &   0.01530 &  -4.1393 & 	0.0072 &  0.0913  &        0.0112   \\
 8909.3483333  &     25.1606 &   0.01743 &  -4.1166 & 	0.0082 &  -0.1150 &       0.0134 \\
 8910.3658396  &     25.3726 &   0.01616 &  -4.1773 & 	0.0086 &  0.0874  &        0.0123 \\
 8912.3642346  &     25.2665 &   0.01417 &  -4.1737 & 	0.0069 &  0.0045  &       0.0087  \\
 8914.3558708  &     25.2970 &   0.01264 &  -4.1760 & 	0.0056 &  0.0245  &       0.0093 \\
 8915.3560174  &     25.2549 &   0.01466 &  -4.1673 & 	0.0069 &  -0.0155 &       0.0098 \\
 8920.3463824  &     25.3692 &   0.01414 &  -4.1867 & 	0.0069 &  0.0521  &       0.0110 \\
 8921.3508119  &     25.3961 &   0.01262 &  -4.1763 & 	0.0056 &  0.1035  &       0.0090   \\ 
 8923.3516166  &     25.3017 &   0.01346 &  -4.1568 & 	0.0060 &  0.0267  &       0.0082  \\
\noalign{\smallskip}
\hline
\noalign{\smallskip}
\end{longtable}

\twocolumn

\clearpage
\end{appendix}

\end{document}